\documentclass[aps,prb,twocolumn,superscriptaddress,footinbib]{revtex4-2}
\usepackage[usenames,dvipsnames]{color}
\usepackage[colorlinks, citecolor={blue}, urlcolor={blue}, linkcolor={blue}]{hyperref}

\usepackage{physics} 
\usepackage{amsmath} 
\usepackage{amssymb} 
\usepackage{bm} 
\usepackage{braket} 
\usepackage{latexsym}

\usepackage[greek,english]{babel}
\newcommand\koppa{\text{\selectlanguage{greek}ϙ}}

\renewcommand{\t}{\text} 

\usepackage{graphicx} 
\usepackage{stackengine} 
\usepackage[caption=false]{subfig} 
\usepackage[inline]{enumitem} 
\setlist[enumerate,1]{label={(\roman*)}} 
\setlist{nolistsep} 
\usepackage{comment} 
\usepackage{lmodern} 

\usepackage{multirow}
\usepackage{hhline}
\usepackage{soul}

\begin{document}

\graphicspath{{./Figures/}}

\newcommand{\QuICS}{Joint Center for Quantum Information and Computer Science, National Institute of Standards and Technology and University of Maryland, College Park, Maryland 20742, USA}
\newcommand{\JQI}{Joint Quantum Institute, National Institute of Standards and Technology and University of Maryland, College Park, Maryland 20742, USA}
\newcommand{\FAU}{Friedrich-Alexander-Universit\"at Erlangen-N\"urnberg (FAU), Department Physik, Staudtstra{\ss}e 7, D-91058 Erlangen, Germany}

\newcommand{\thetitle}{Unconventional entanglement scaling and quantum criticality in the long-range spin-one Heisenberg chain with single-ion anisotropy}

\title{\thetitle}
\author{Patrick Adelhardt}
\affiliation{\FAU}
\affiliation{\QuICS}
\affiliation{\JQI}
\author{Sean R.~Muleady}
\affiliation{\QuICS}
\affiliation{\JQI}
\author{Kai P.~Schmidt} 
\affiliation{\FAU}
\author{Alexey V.~Gorshkov}
\affiliation{\QuICS}
\affiliation{\JQI}

\date{\today}

\begin{abstract} 
Long-range interactions can fundamentally reshape the low-energy properties of low-dimensional quantum matter, altering both continuous symmetry breaking and topological phenomena. However, their impact on the quantum criticality separating these regimes remains poorly understood. We determine the ground-state phase diagram and critical properties of the spin-one Heisenberg chain with single-ion anisotropy and staggered antiferromagnetic power-law interactions, using matrix-product state (MPS) calculations complemented by high-order series expansions (pCUT+MC). Such long-range, non-frustrated interactions circumvent the Hohenberg-Mermin-Wagner theorem, thereby stabilizing continuous symmetry breaking (CSB) phases in direct competition with the Haldane phase. We map out the resulting phase diagram and analyze the entanglement entropy scaling behavior in the U(1) and SU(2) CSB phases, finding logarithmic corrections beyond the short-range, universal contributions expected from linearly dispersed Goldstone modes. We further characterize all critical boundaries through finite-size scaling of either the entanglement entropy or the staggered magnetization. In particular, the large-D-to-U(1)-CSB transition exhibits unconventional, continuously varying critical exponents as a function of the long-range decay exponent with a strong dependence on the imposed boundary conditions leading to distinct finite-size scalings for sufficiently long-range potentials. Remarkably, the Haldane-to-U(1)-CSB transition likewise displays unconventional quantum criticality with distinct continuously varying critical exponents. Our work positions this model as a target for near-term atomic platforms with tunable long-range couplings and exhibiting natural single-ion anisotropy, offering a minimal playground for exploring the interplay between long-range interactions, continuous symmetry breaking, and topology.
\end{abstract}

\maketitle

\section{Introduction}
\label{sec:intro}

The spin-one Heisenberg chain is a paradigmatic model for the interplay between topology, symmetry, and quantum fluctuations. With nearest-neighbor antiferromagnetic couplings, it hosts not only trivial and antiferromagnetically ordered phases, but also the celebrated Haldane phase: a symmetry-protected phase predicted by the Haldane conjecture~\cite{haldane_continuum_1983,haldane_nonlinear_1983,affleck_proof_1986,affleck_critical_1987,chiu_classification_2016,haldane_nobel_2017}. In particular, this phase is a manifestation of the fact that integer-spin systems host a gapped spectrum and localized, stable edge excitations, in stark contrast to their gapless, half-integer spin counterparts well-described by spin-wave theory~\cite{anderson_approximate_1952}. A broad issue concerns how this structure is altered by the addition of long-range couplings, which can evade the constraints of the Hohenberg-Mermin-Wagner theorem ~\cite{Mermin1966,Hohenberg1967}, thereby allowing spontaneous breaking of continuous symmetries in low dimensions. This raises a central question regarding the fate of the Haldane phase in the presence of long-range interactions, and its competition with continuous symmetry breaking~\cite{Gong2016a,Gong2016b}.

Such issues are also timely from the standpoint of quantum simulation. In recent years, significant progress in the realization and control of atomic and molecular platforms for quantum science---including trapped ions~\cite{blatt_quantum_2012,monroe_programmable_2021}, Rydberg atom arrays~\cite{browaeys_many-body_2020,kaufman_quantum_2021}, dipolar magnets~\cite{chomaz_dipolar_2023}, and polar molecules~\cite{bohn_cold_2017}---has provided increasing access to the physics of long-range interacting systems~\cite{defenu_long-range_2023}. For instance, these systems have led to the recent observation of entanglement generation~\cite{bornet_scalable_2023,franke_quantum-enhanced_2023,eckner_realizing_2023}, non-local transport properties~\cite{richerme_non-local_2014,jurcevic_quasiparticle_2014,joshi_observing_2022}, and continuous-symmetry breaking~\cite{feng_continuous_2023,chen_continuous_2023}. While such systems naturally host interactions that monotonically decay algebraically with distance between particles, recent experiments in trapped ion and cavity-based neutral atom platforms have also demonstrated the ability to engineer not only tunable interaction ranges, but also non-trivial sign structures~\cite{periwal_programmable_2021,katz_floquet_2025}. Of particular relevance here, several proposals and experiments have focused on realizing spin-one Heisenberg-type models with tunable interactions in dipolar species~\cite{Chung2021,Sompet2022}, trapped ions~\cite{Cohen2014,Cohen2015,Senko2015}, and Rydberg platforms~\cite{Brechtelsbauer2025,Moegerle2025}, typically with the addition of single-ion anisotropy terms arising naturally as quadratic Zeeman terms.

While some numerical studies have explored long-range Heisenberg and XXZ interactions in $S=1$ systems~\cite{Gong2016a, Gong2016b, Ren2020, Kawasaki2025}, including cases with ferromagnetic~\cite{Gong2016b,Kawasaki2025} or frustrated antiferromagnetic couplings~\cite{Gong2016a,Gong2016b} and single-ion anisotropy~\cite{Ren2020}, the interplay with staggered (unfrustrated) antiferromagnetic long-range interactions remains largely unexplored. In contrast to frustrated long-range interactions, which tend to suppress conventional magnetic ordering~\cite{Gong2016a, Gong2016b,Ren2020}, unfrustrated long-range couplings allow one to circumvent the consequences of the Hohenberg-Mermin-Wagner theorem~\cite{Hohenberg1967,Mermin1966,Pitaevskii1991,Bruno2001} in one-dimensional systems and ensure stable ordered phases with breaking of both continuous SU(2) and U(1) symmetries, thereby enabling a direct investigation of their interplay with the Haldane phase.

Here, we characterize the ground state and critical properties of the spin-one Heisenberg model, with staggered, long-range interactions and single-ion anisotropy.
Employing large-scale matrix product state~\cite{Schollwoeck2011,Fishman2022} calculations using the density matrix renormalization group~\cite{white_density_1992,White1993}, as well as high-order series expansions (pCUT+MC)~\cite{Fey2019, Adelhardt2024}, we map out the relevant ground-state phase diagram in terms of both the correlation functions and the scaling of the entanglement entropy with system size. Our results reveal several unconventional features. Most notably, the transition between the disordered ``large-D'' and U(1) symmetry-broken phases exhibits continuously varying critical exponents as a function of the interaction decay exponent and boundary-dependent finite-size scalings for sufficiently strong long-range potentials. The transition from the Haldane to the U(1) symmetry-broken phase demonstrates a distinct, continuously varying dependence on the model parameters. In all, our work establishes this model as a rich playground for exploring the complex interplay of long-range interactions, topology, and continuous symmetry breaking. These results provide a direct route for experiments harnessing controllable spin-spin interactions in higher spin systems to explore unconventional critical behavior in near-term quantum simulators.

The remainder of this paper is organized as follows. In Sec.~\ref{sec:model}, we introduce the model Hamiltonian and briefly outline the methodological aspects of this work. In Sec.~\ref{sec:phase_diagram}, we present and discuss the ground-state phase diagram, followed in Sec.~\ref{sec:phase_transitions} by a detailed discussion of all second-order quantum phase transitions. Finally, Sec.~\ref{sec:conclusion} summarizes our results and presents our conclusions.

\section{Model and Methods}
\label{sec:model}

We study the following one-dimensional spin-one Heisenberg Hamiltonian with single-ion anisotropy and staggered (non-frustrated) antiferromagnetic long-range interactions:
\begin{equation}
    H = D \sum_i \left(S^z_i\right)^2 + \frac12 \sum_{i\neq j} J(j-i) \bm{S}_i \cdot\bm{S}_j\,.
    \label{eq:ham}
\end{equation}
Here, $\bm{S}_i=(S_i^x,S_i^y,S_i^z)^T$ where $S_i^{\sigma}$ are spin-one operators at a site $i$ with $\sigma\in\{x,y,z\}$. The strength of the single-ion anisotropy is given by $D$, and the long-range interaction strength is denoted as $J(j-i)$, which depends on the distance between the interacting sites $i$ and $j$. We denote the eigenstates of the spin-one operator $S_i^{z}$ as $\ket{\kappa_i}$ with the eigenvalue $\kappa_i \in\{0,\pm 1\}$. The model with nearest-neighbor interactions (\mbox{$J(j-i)\propto \delta_{j,i+1}$}) and its generalization to XXZ interactions has been thoroughly investigated in recent decades~\cite{Botet1983,Sakai1990,Tonegawa1996,Chen2000,Chen2003,CamposVenuti2006,Tzeng2008a,Tzeng2008b,Ueda2008,Albuquerque2009,Adelhardt2017}.  

Here, we consider the case of staggered (non-frustrated) antiferromagnetic interactions
\begin{equation}
    J(j-i) \equiv \frac{(-1)^{j-i+1}}{|j-i|^{\alpha}} \,,
    \label{eq:stag_lr}
\end{equation}
where the long-range decay exponent $\alpha$ controls the power-law falloff with distance. In the limit $\alpha=\infty$, the system exhibits antiferromagnetic nearest-neighbor interactions. For $\alpha\le 1$, the system becomes superextensive, approaching a uniform staggered all-to-all coupling in the limiting case $\alpha=0$.  In contrast to uniformly decaying long-range interactions, the unfrustrated nature of the couplings \eqref{eq:stag_lr} circumvents the Hohenberg-Mermin-Wagner theorem~\cite{Hohenberg1967,Mermin1966,Pitaevskii1991,Bruno2001}, allowing spontaneous breaking of continuous symmetries. This enables the realization of an SU(2) continuous symmetry-breaking (CSB) phase at $D=0$ and a U(1) CSB phase for $D>0$. We note that such interactions have been previously considered in the absence of the single-ion anisotropy term~\cite{Gong2016a}; a related study has also considered uniformly decaying power-law interactions with a generalized staggered single-ion anisotropy~\cite{Ren2020}.

For $D=+\infty$, the ground state of the nearest-neighbor model is given by the product state $\prod_i \ket{0_i}$ as $\ket{0}$ is the lowest lying state of a single $D (S^z)^2$ term. In the thermodynamic limit, the phase at large $D$ is still adiabatically connected to this product state and is referred to as the ``large-D'' phase. In the opposite limit $D\rightarrow -\infty$, the system exhibits a staggered antiferromagnetic order in the z-direction ($\t{Z}_2$ AF), since a negative $D$ flips the energy spectrum of the single-ion anisotropy, so that $\ket{\pm 1}$ become the lowest lying states. For $D=0$, we are left with the spin-one Heisenberg chain exhibiting a gapped ground state, the so-called Haldane phase~\cite{Haldane1983a,Haldane1983b}. 

In this work, we use matrix product states~\cite{Schollwoeck2011,Fishman2022} to calculate the entanglement entropy $S_{\t{VN}}$, the staggered longitudinal magnetization $M_{z}$ and transverse magnetization $M_{\perp}$ as well as various spin-spin correlations for finite system sizes up to $L=340$. Finite-size scaling~\cite{Fisher1972, Binder1987, Cardy1988Book, Koziol2021,Langheld2022, Stanley1999} allows us to determine the ground-state phase diagram and the properties of the critical lines. To efficiently represent the algebraically decaying long-range interactions $f(r)\equiv r^{-\alpha}$ in the MPS formalism, we approximate the power law with a sum of $N$ exponential functions by $\tilde{f}(r) \equiv \sum_{l=1}^{N} a^{\phantom{r}}_l b_l^{r-1}$~\cite{Crosswhite2008,Pirvu2010}, minimizing the cost function
\begin{equation}
    \vartheta(\{a_l\}, \{b_l\}) = \sum_{r=1}^L \left(f(r) - \tilde{f}(r)\right)^2
    \label{eq:exp_approx}
\end{equation}
and keeping the error below $\varepsilon \leq 10^{-9}$ with up to $N=49$ exponential terms for the largest system size $L=340$. To find the ground state of the system, we employ the density matrix renormalization group (DMRG) algorithm~\cite{white_density_1992,White1993}. We use a maximal bond dimension between $\chi_{\t{max}}=500$ and $\chi_{\t{max}}=2000$. Throughout, we restrict ourselves to the $m_z=0$ symmetry sector when not stated otherwise, where $m_z$ is the eigenvalue of the total spin operator $S^z_{\t{tot}} = \sum_i S_i^z$.

We complement the MPS approach with high-order series expansions, expanding from the non-interacting limit $D=\infty$ in powers of the perturbation parameter $\lambda = 1/D$. To handle long-range interactions, we use the pCUT+MC approach based on a linked-cluster expansion via a full graph decomposition~\cite{Adelhardt2024}. This approach combines the method of perturbative continuous unitary transformations (pCUT)~\cite{Knetter2000,Knetter2003} for the calculation of perturbative contributions on so-called white graphs~\cite{Coester2015} with a Monte Carlo (MC) summation algorithm to embed the white-graph contributions in the long-range chain to determine the physical quantities of interest in the thermodynamic limit~\cite{Fey2016,Fey2019, Koziol2019, Adelhardt2020,Adelhardt2023, Adelhardt2024, Adelhardt2025}. We calculate the elementary excitation gap and the corresponding spectral weight to orders 10 and 9 in the perturbation parameter $\lambda$, respectively. For technical details, we refer the reader to Ref.~\cite{Adelhardt2024}.

\section{Quantum phase diagram}
\label{sec:phase_diagram}

\begin{figure}[t!]
	\centering
	\includegraphics[width=1.\columnwidth]{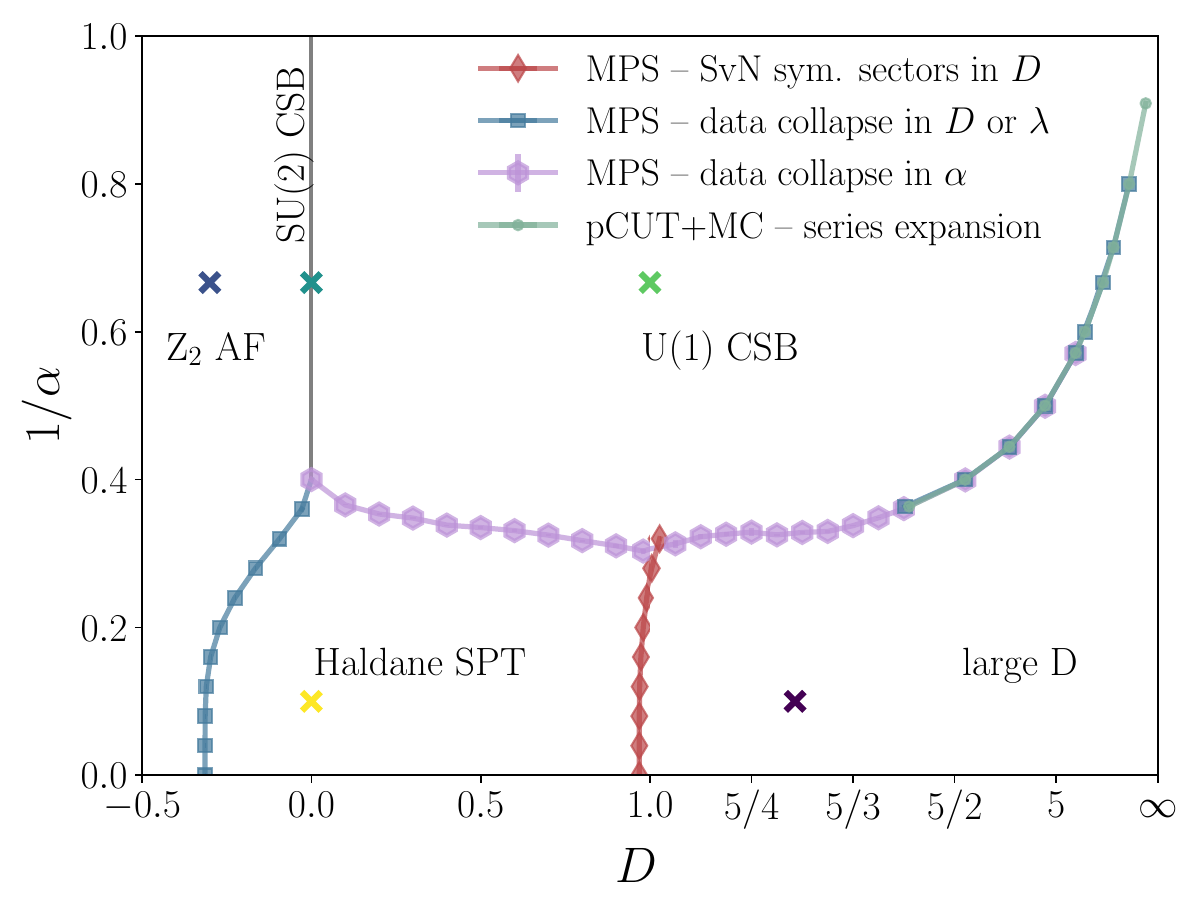}
	\caption{Quantum phase diagram as a function of the single-ion anisotropy $D$ and long-range decay exponent $\alpha$. The $D$-axis is linear for $D\le 1$, while for $D>1$ it is uniformly spaced in $1/D$. Five phases appear: the z-antiferromagnetic ($\t{Z}_2$ AF) phase, the  Haldane symmetry protected topological (Haldane SPT) phase, the large-D phase, and the U(1) and SU(2) continuous-symmetry-breaking (CSB) phases. Diagonal crosses mark the points where correlations are evaluated in Fig.~\ref{fig:fig2}. Transition lines are obtained from the indicated methods and are distinguished by color and marker style. The gray line at $D=0$ is a high-symmetry line with SU(2) CSB.}
	\label{fig:fig1}
\end{figure}

In this section, we investigate the ground-state phase diagram, as a function of the single-ion anisotropy $D$ and the long-range decay exponent $\alpha$, as depicted in Fig.~\ref{fig:fig1}. 
As we described above, the Hamiltonian \eqref{eq:ham} realizes an antiferromagnetic phase (`$\t{Z}_2$ AF') for sufficiently negative values of $D$, a Haldane phase (`Haldane SPT') for small $D$ around the origin, and a trivial disordered phase (`large D') for large $D$ values. The Haldane phase falls into the category of one-dimensional topological phases referred to as symmetry-protected topological phases (SPTs)~\cite{Ejima2015,Pollmann2010,Chen2011a, Chen2011b,Pollmann2012,Liu2014}.

As we make the interactions stronger by decreasing the value of the decay exponent $\alpha$, we induce antiferromagnetic ordering by the non-frustrated nature of the staggered long-range interactions, leading to continuous symmetry breaking (CSB). For $D=0$, the Hamiltonian in the thermodynamic limit is invariant under global SU(2) spin rotations, i.\,e.\,, $[H,\bm{S}_{\t{tot}}]=0$, where $\bm{S}_{\t{tot}}=\sum_i \bm{S}_i$. Thus, the ordered ground state spontaneously breaks the full SU(2) symmetry. For a finite anisotropy ($D\neq 0$), the anisotropy explicitly reduces the symmetry to global U(1) rotations about the z-axis, generated by $S^z_{\t{tot}}=\sum_i S^z_i$, i.\,e.\,, $[H,S^z_{\t{tot}}]=0$ still holds, but $[H,S^{x/y}_{\t{tot}}]\neq 0$. In this case, sufficiently strong long-range interactions lead to spontaneous breaking of this remaining U(1) symmetry. 
A compact overview of the phases and their main characteristic properties is provided in Table~\ref{tab:phase_summary}. 

\begin{table}[t]
\caption{Summary of the characteristic properties of the phases appearing in the phase diagram. We indicate whether each phase is gapped or gapless, the symmetry that is broken (if any), the nature of the order parameter, and the corresponding decay of correlations. We use the abbreviation ``LRO'' for long-range order and ``SPT'' for symmetry-protected topological order. The notation $\langle M_{x,y,z} \rangle$ denotes the staggered antiferromagnetic order parameters in the $x$, $y$, and $z$ spin components, respectively.}
\label{tab:phase_summary}
\begin{ruledtabular}
\begin{tabular}{lcccc}
Phase & Gap & Broken sym. & Order & Correlations \\
\hline
Haldane
& Yes & None & SPT & Exponential \\

Large-$D$
& Yes & None & None & Exponential \\

$Z_2$ AFM
& Yes & $Z_2$ & $\langle M_z \rangle \neq 0$ & LRO  \\

U(1) CSB
& No & U(1) & $\langle M_{x,y}\rangle \neq 0$ & LRO  \\

SU(2) CSB
& No & SU(2) & $\langle M^{x,y,z}\rangle \neq 0$ & LRO \\
\end{tabular}
\end{ruledtabular}
\end{table}

\subsection{Spin-spin correlations}

\begin{figure}[t]
	\centering
	\includegraphics[width=1.\columnwidth]{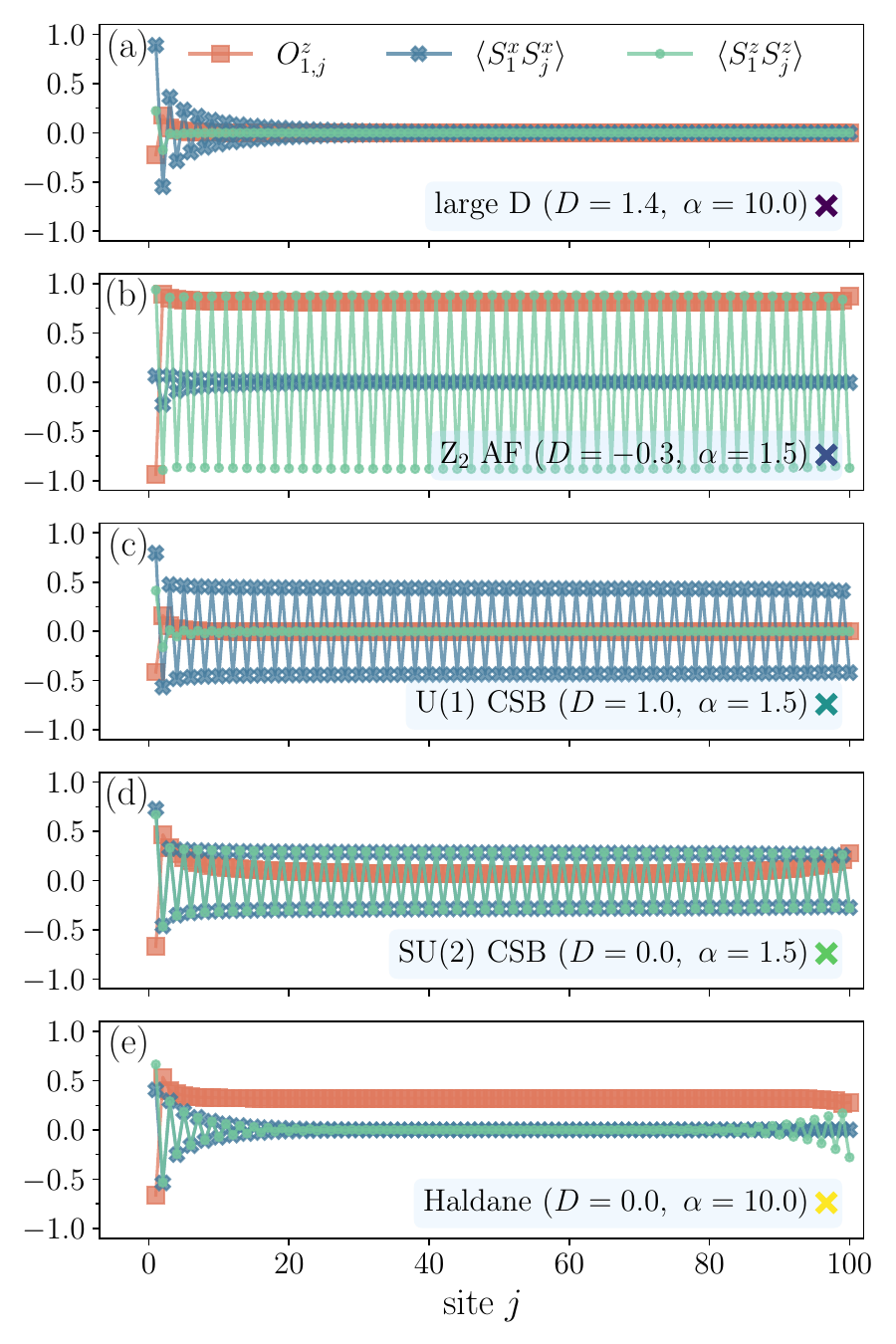}
	\caption{Spin-spin correlations $\langle S^{\sigma}_1 S^{\sigma}_j\rangle$ for the $\sigma=x, z$ spin components, as well as the string-order correlator $O^z_{1,j}$ between site $1$ and an arbitrary site $j$. The trivial large-D phase (a) is characterized by exponentially decaying correlations for all quantities. The $\t{Z}_2$ AF phase (b) exhibits finite staggered spin correlations in the $z$ component, the U(1) CSB phase (c) in the $x$ (and $y$) component, and the SU(2) CSB phase (d) in all three components. The Haldane phase (e) shows exponentially decaying spin correlations but a finite string-order correlator. Results are shown for a chain with $L=100$ for all five quantum phases of the model. Diagonal crosses in the labels indicate cross-references to the phase diagram in Fig.~\ref{fig:fig1}, marking the parameter points at which the correlations were evaluated.}
	\label{fig:fig2}
\end{figure}

To unambiguously characterize  the aforementioned quantum phases, we probe the spin-spin correlations $\langle S^\sigma_1 S^\sigma_j \rangle$ of the spin components $\sigma$ between the first site and an arbitrary site $j$, as well as the non-local string order correlator 
\begin{equation}
    O^z_{1,j} = -S_1^z \left(\prod_{i=2}^{j-1}e^{-\t{i}\pi S_i^z}\right) S_j^z\,,
\end{equation}
which we can use to locate the Haldane phase~\cite{Nijs1989}. Note that $\langle S^x_1 S^x_j \rangle=\langle S^y_1 S^y_j \rangle$ always holds due to the U(1) symmetry of the Hamiltonian. In Fig.~\ref{fig:fig2}, we show the spin-spin correlations and the string order correlator within each phase. 
See also Table~\ref{tab:phase_summary} for the expected behavior of correlation decay in each phase.

We observe the expected exponentially decaying spin-spin correlations in the disordered large-D and Haldane phases. The two phases can be unambiguously distinguished by the string order correlator which vanishes exponentially in the trivial large-D phase but remains finite in the Haldane phase. The $\t{Z}_2$ AF phase is characterized by finite staggered correlations in the z component with values close to $\approx\pm 1$, exponentially vanishing correlations in the transverse direction, and a trivially finite string-order correlator. The CSB phases exhibit staggered correlations distributed over two (three) components with values $\approx \pm 0.43$ ($\approx \pm 0.28$) for U(1) (SU(2)) CSB. These values are close to the classical Néel limits $1/2$ ($1/3$) for $S=1$, indicating a nearly saturated ordered moment. The proximity to the classical values reflects the suppression of long-wavelength quantum fluctuations by the unfrustrated long-range interactions, stabilizing the ordered state.

\subsection{Entanglement scaling}
\label{sec:ee_scaling}

We study the scaling behavior of the von Neumann entanglement entropy $S_{\t{VN}}$ in the various quantum phases as a function of the decay exponent $\alpha$. In most settings, the entanglement entropy obeys a leading area-law scaling with linear system size $L$,
\begin{equation}
    S_{\t{VN}} \sim a\,L^{\,d-1} + b \ln L + c,
    \label{eq:area_law}
\end{equation}
where $d$ is the spatial dimension, $a$ the coefficient of the dominant area-law contribution, $b$ the prefactor of the logarithmic correction, and $c$ a constant offset~\cite{Laflorencie2016}. Note, for one-dimensional systems ($d=1$), the leading area-law contribution is constant and is therefore absorbed into the offset $c$. The origin of these contributions often depend directly on the nature of the phase. Of particular relevance for the model here are the following cases. 
In gapped ordered phases resulting from discrete symmetry breaking, the entanglement entropy obeys an area law with a constant correction given by the ground-state degeneracy~\cite{Laflorencie2016}.
In gapless phases associated with continuous symmetry breaking, $b=N_{\t{G}}/2$ is directly linked to the number of Goldstone modes $N_{\t{G}}$~\cite{Laflorencie2016}. Finally, in one-dimensional critical phases described by conformal field theory, the logarithmic term is governed by an effective central charge $c_{\mathrm{eff}}$~\cite{Calabrese2004, Laflorencie2016}. 

We calculate the entanglement entropy for \mbox{$D\in\{-0.5,0,0.5,1.25\}$} for various values of $\alpha$ and use a systematic fitting and extrapolation procedure to determine $b$ and $c$ as described in Appendix \ref{app:ee_scaling}. 
\begin{figure}[t]
	\centering
	\includegraphics[width=1.\columnwidth]{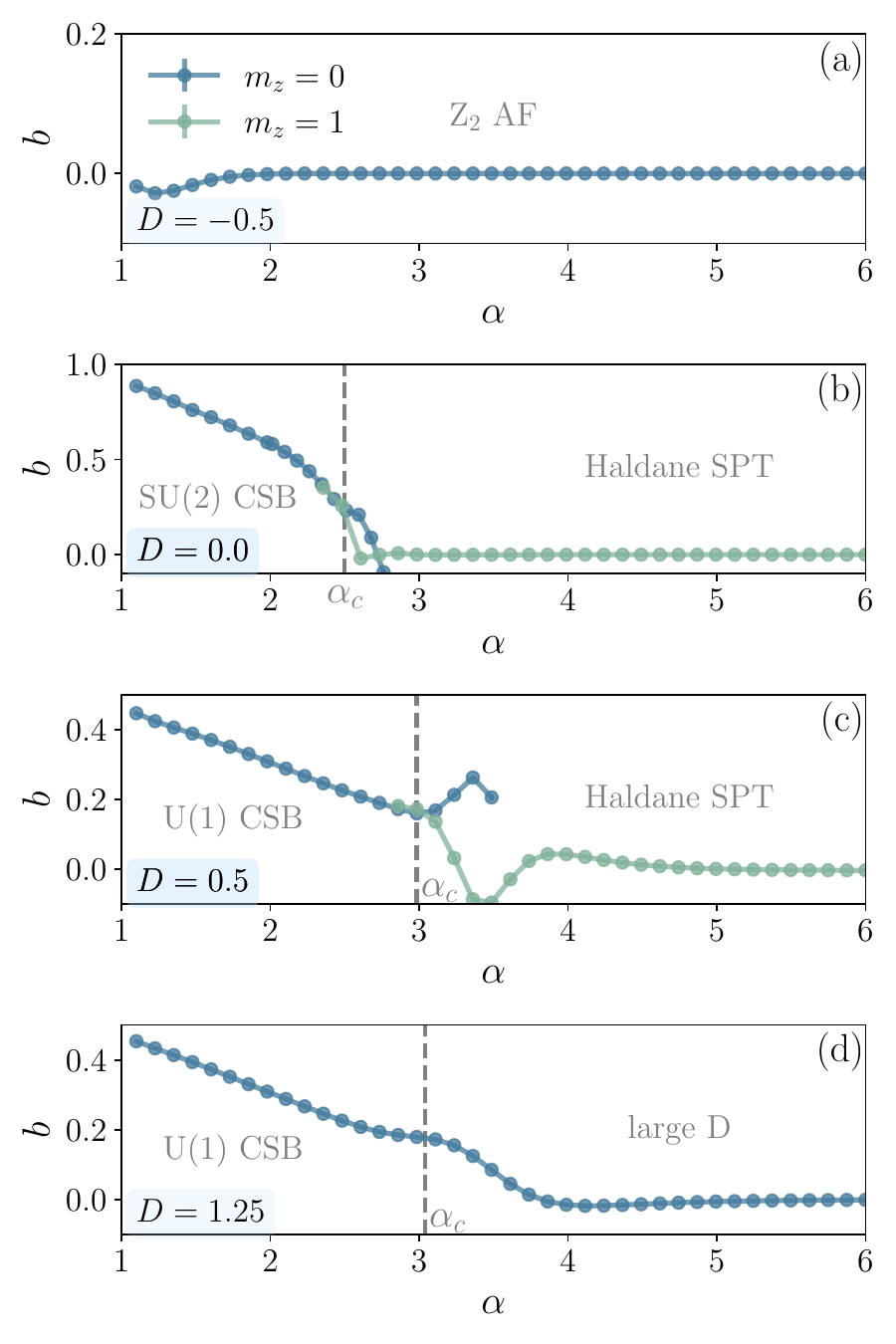}
	\caption{Logarithmic coefficient $b$ of the entanglement entropy as a function of the decay exponent $\alpha$ for \mbox{$D = -0.5$, $0$, $0.5$, and $1.25$} in panels (a), (b), (c), and (d), respectively. Data points include errorbars from the extrapolation procedure. The ground state in the $m_z = 0$ sector is used to compute the entanglement entropy, except in the Haldane phase, where the $m_z = 1$ ground state is taken to avoid the residual $\ln 2$ contribution from edge-state entanglement. Fitting the entanglement entropy to $S_{\mathrm{VN}} = b \ln L + c$ yields the logarithmic coefficient $b$, which displays a smoothly varying, unconventional scaling within the CSB phases (panels (b), (c), and (d)), while remaining constant in all other phases. For $\alpha \to 1$, the coefficients approach $b = N_{\t{G}}/2$, determined by the number of Goldstone modes $N_{\t{G}}$, where $N_{\t{G}}=1$ for U(1) CSB in (c) and (d) and $N_{\t{G}}=2$ for SU(2) CSB in (b). Gray dashed lines indicate critical points in the thermodynamic limit determined by the data-collapse and extrapolation procedure in Appendix~\ref{app:data_collapse}.}
	\label{fig:fig3}
\end{figure}
We show the results for $b$ as a function of the decay exponent $\alpha$ in Fig.~\ref{fig:fig3}. A supplementary discussion of the constant term $c$ can be found in Appendix \ref{app:ee_scaling}. 

As expected, for the $\t{Z}_2$ AF phase (\mbox{$D=-0.5$}, Fig.~\ref{fig:fig3}\,(a)), we observe a constant behavior of the entanglement entropy such that $b\approx 0$ for all $\alpha$ values due to the gapped nature of the phase. Interestingly, for $\alpha\lesssim 2$, the coefficient $b$ becomes negative. We attribute this to finite-size effects: within the accessible system sizes, the entanglement entropy slightly decreases with increasing $L$, leading to a small negative slope in the logarithmic fit, while the entropy is expected to approach a positive value in the thermodynamic limit (c.\,f.~Fig.~\ref{fig:figA2} in Appendix~\ref{app:ee_scaling}).

The behavior in the gapped Haldane phase is more subtle, owing to its nature as a topologically protected phase. This phase is characterized by the existence of effective spin-1/2 degrees of freedom at the edges of the chain. These edge spins are governed by the effective Hamiltonian \mbox{$H_{\rm edge} = J_{\rm edge}\bm{S}_1\cdot\bm{S}_N$}, for spin-1/2 operators $\bm{S}_i$ and effective coupling $J_{\rm edge}>0$ that decays exponentially with the system size~\cite{Hu2011}. In the ground state with $m_z=0$, the edge spins form a singlet state $\ket{s}$ with non-vanishing correlations that span the extent of the chain. This produces a constant $\ln(2)$ contribution to the entanglement entropy arising from edge-state entanglement. In practice, this contribution persists in our simulations as long as the effective edge-spin coupling $J_{\rm edge}$ exceeds the numerical tolerance. For finite long-range interactions ($\alpha <\infty$), the edge spins couple directly, resulting in a power-law decay of $J_{\rm edge}$ rather than the exponential decay observed in the nearest-neighbor model, thereby stabilizing the edge-spin entanglement.
This issue can be overcome by working in the $m_z=1$ sector, where the ground state of the edge spins is the spin-polarized triplet state. This state is a product state with no inter-spin correlations, allowing us to cleanly observe the bulk behavior of the entanglement entropy at small system sizes.

In panels (b) and (c) of Fig.~\ref{fig:fig3} we plot the entanglement entropy using the ground state in the \mbox{$m_z=1$} sector for the Haldane phase, and the ground state in the $m_z=0$ sector for small $\alpha$ values in the U(1) and SU(2) CSB phases. For large enough $D\gtrsim 1$ (see Fig.~\ref{fig:fig3}\,(d)), we return to calculating the ground state in the $m_z=0$ sector for all $\alpha$ values as the large-D phase emerges. 

When comparing the behavior of $D\in \{0,0.5,1.25\}$, we find that the logarithmic coefficient is zero in the Haldane and large-D phases for \mbox{$\alpha\gtrsim \alpha_c$} and increases smoothly in the CSB phases for \mbox{$\alpha\lesssim \alpha_c$}, where $\alpha_c$ is the critical value of the decay exponent $\alpha$. The distinct scaling behavior of the entanglement entropy in the CSB phases was found relatively recently in the long-range spin-1/2 Heisenberg chain~\cite{Frerot2017,Zhao2025}. This $\alpha$-dependent behavior of the logarithmic correction goes beyond the scenario of CSB in nearest-neighbor antiferromagnetic models where Goldstone modes exhibit a linear dispersion, and reflects the modification of the low-energy behavior by the long-range interactions, which leads to a sublinear dispersion that continuously varies with $\alpha$~\cite{Diessel2023,Song2023,Adelhardt2025}.

We also observe that the scaling behavior in the SU(2) CSB phase differs from that in the U(1) CSB phase. We can understand this difference as the result of the number of Goldstone modes $N_{\t{G}}$ in the limiting case $\alpha\rightarrow 1$, where we expect a logarithmic contribution with $b=N_{\t{G}}/2$~\cite{Vidal2007,Laflorencie2016,Frerot2017,Misguich2017,Zhao2025}.
At $\alpha=0$, where the system exhibits all-to-all staggered interactions, the system is well-described by a collective spin-$S$ model, with $S=N/2$, where CSB is the result of a canonical tower of states structure~\cite{Frerot2017}. For SU(2) symmetry breaking it was shown for the Lieb-Mattis model~\cite{Lieb1962} that $b(\alpha=0) = 1$~\cite{Vidal2007}, and for U(1) symmetry breaking the Lipkin–Meshkov–Glick model~\cite{Lipkin1965,Meshkov1965} yields $b(\alpha=0) = 1/2$~\cite{Vidal2007}. These results are expected to remain valid throughout the super-extensive regime $0\le \alpha\le 1$~\cite{Zhao2025}. We find $b(\alpha=1)\approx 0.92$ for SU(2) CSB and $b(\alpha=1)\approx 0.47$ for U(1) CSB, consistent with the fact that $N_{\t{G}}=2$ for SU(2) CSB and $N_{\t{G}}=1$ for U(1) CSB. We attribute the slight underestimation of these values to errors arising from the restricted bond dimension at small $\alpha$, where the formation of long-range correlations drastically increases the computational cost of MPS.

Finally, we determine the coefficient $b$ at the critical point in Fig.~\ref{fig:fig3}\,(b), (c), and (d). Analogously to Ref.~\cite{Zhao2025}, we expect an abrupt change of $b$ from zero to a finite value at the quantum phase transition. Here, we observe a smooth crossover with a plateau close to the expected transition.
We attribute this smooth behavior of $b$ across the transition to finite-size effects which can mask sharp features expected at the phase transition in the thermodynamic limit. At the critical decay exponent $\alpha_c$---determined via data collapses and the rigorous extrapolation procedure described in Appendix \ref{app:data_collapse}--- we determine the effective critical logarithmic coefficients $b_c=b(\alpha_c)\approx 0.24, 0.16, 0.18$ for $D=0,0.5,1.25$, respectively. These values show a noticeable variation with the anisotropy $D$. Note that the critical value $b_c=0.275(9)$ found in the long-range spin-1/2 Heisenberg chain with SU(2) CSB~\cite{Zhao2025} also differs from the value found in our model at $D=0$. At $D=0.5$, we observe pronounced features in both symmetry sectors---$m_z=0$ and $m_z=1$---next to the critical point $\alpha_c$, at slightly larger values of the decay exponent. Such features can arise from stronger scaling corrections in the critical region. Moreover, the shift is consistent with the notion of a finite-size pseudo-critical point, where extrema of observables are shifted relative to the thermodynamic transition~\cite{Binder1987}.

\section{Quantum phase transitions}
\label{sec:phase_transitions}

In this section, we investigate the quantum phase transition lines obtained in Fig.~\ref{fig:fig1} by analyzing their finite-size behavior. We focus on the transitions identified as second order, while the high-symmetry line at $D=0$ between Z$_2$ AF and U(1) CSB phases is first order and not considered further. We begin with the Gaussian transition between the Haldane and large-D phases, where the topological character of the transition requires an entanglement entropy scaling analysis in different symmetry sectors. We then examine the Ising transition between the Haldane and the $\t{Z}_2$ AF phase, using data collapses of the magnetization curves for various system sizes to establish the critical behavior and to benchmark our approach. Finally, we analyze the various CSB transitions, again employing data collapses and comparing the resulting behavior across distinct parameter regimes.

\subsection{Gaussian transition}

We start by analyzing the Gaussian transition between the Haldane and large-D phases. As neither the Haldane phase nor the large-D phase breaks any global symmetry of the Hamiltonian, the transition cannot be described within the Landau-Ginzburg-Wilson framework~\cite{Wilson1974, Wilson1983}. Instead, this corresponds to a topological quantum phase transition. The low-energy field theory of the spin-one chain with single-ion anisotropy can be mapped, after integrating out massive modes, to an effective Sine-Gordon model,
\begin{equation}
    \mathcal{L} = \frac{1}{2}(\partial_{\mu}\phi)^2 
        + g \cos(\beta\phi),
\end{equation}
where the first term describes a Gaussian free boson field~\cite{Chen2000,Degli2003,Gogolin2004Book}. In the renormalization group sense, the cosine interaction is relevant for $\beta^2 < 8\pi$~\cite{Gogolin2004Book}, in which case the coupling $g$ flows to strong coupling, the field $\phi$ becomes pinned and a spectral gap opens. This occurs in both the Haldane and large-D phases. At the critical point $D_c$, the coupling becomes marginal, so the effective theory reduces to a gapless Gaussian model with central charge $c_{\mathrm{eff}} = 1$, which defines the Gaussian transition~\cite{Gogolin2004Book, Degli2003}. In Sec.~\ref{sec:ee_scaling}, we discussed the drop in the entanglement entropy upon increasing the system size. A similar behavior can be observed in our calculations when increasing the anisotropy $D$ (see also Ref.~\cite{Hu2011}).

\begin{figure}[t]
	\centering
	\includegraphics[width=1.\columnwidth]{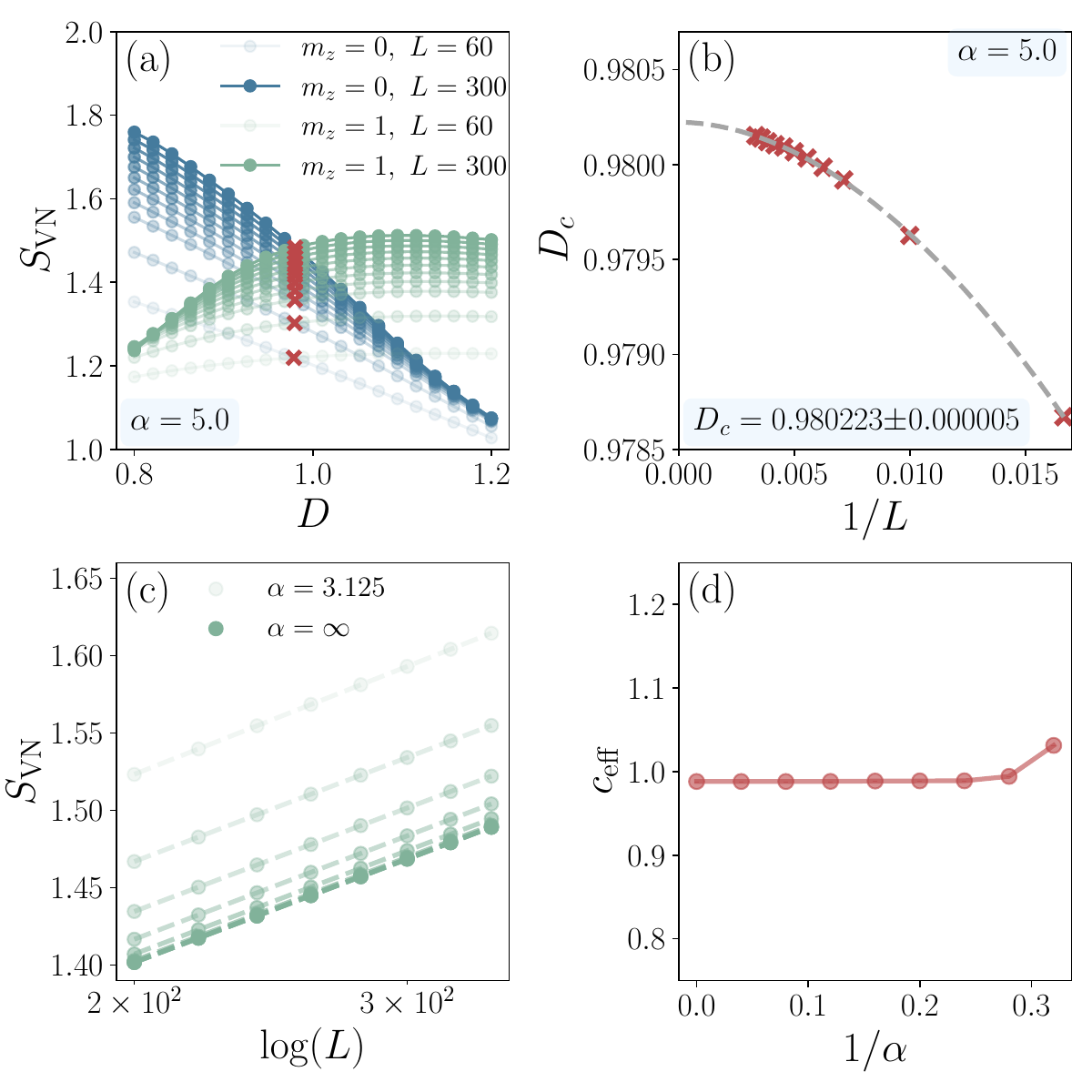}
	\caption{Analysis of the Gaussian transition between the Haldane and large-D phases exploiting the crossing points of entanglement entropy $S_{\t{VN}}$ from different symmetry sectors with eigenvalues $m_z\in\{0,1\}$ of the total spin operator $S^z_{\t{tot}}=\sum_i S^z_i$. (a) Crossing points of the entanglement entropy $S_{\mathrm{VN}}$ between the ground states in the $m_z = 0$ and $m_z = 1$ sectors for various system sizes $L$ from $L=60$ to $L=300$ identify the maximum of the entanglement entropy and thus provide an estimate of the critical point. (b) The crossing points plotted versus the inverse system size $1/L$ and fitted with $D_c(L) = D_c + a L^{-b}$ yield the critical point $D_c$ in the thermodynamic limit (here, we find $D_c=0.980223(5)$ for $\alpha=5$). (c) At the critical point, the entanglement entropy scales as $(c_{\t{eff}}/6) \,\ln L$, consistent with a conformal field theory description of the Gaussian transition. The effective central charge $c_{\mathrm{eff}}$ is extracted from the slope of the logarithmic scaling. (d) The central charge $c_{\t{eff}}$ is unity as a function of the decay exponent $\alpha$, with deviations appearing as the Gaussian transition approaches another critical line towards the U(1) CSB phase.}
\label{fig:fig4}
\end{figure}

Therefore, we follow the ideas in Ref.~\cite{Hu2011} and calculate the entanglement entropy in both symmetry sectors $m_z=0$ and $m_z=1$ around the quantum phase transition. Within the symmetry sector $m_z=1$, we can cleanly access the bulk ground-state properties with unentangled edge spins of the Haldane phase at small system sizes for $D<D_c$, while the $m_z=0$ sector gives access to the bulk ground-state properties of the large-D phase for $D>D_c$. As in Ref.~\cite{Hu2011}, the crossing points of both entanglement entropies then determine the location of the maximum of the entanglement entropy at the quantum phase transition, which allows us to determine the critical point very accurately, since the maximum is almost invariant with respect to the system size $L$ as shown in Fig.~\ref{fig:fig4}. 
To extract the critical point in the thermodynamic limit, we fit the crossing points to the function $D_c(L) = D_c + aL^{-b}$, where $a$, $b$, and $D_c$ are free fit parameters. 
We can now calculate the entanglement entropy exactly at the critical point for various system sizes.
Under open boundary conditions and a half-chain bipartition ($\ell = L/2$), the expected scaling of the entanglement entropy at criticality  from conformal field theory is
\begin{equation}
    S_{\rm VN}(L) = \frac{c_{\rm eff}}{6}\ln(L) + \mathrm{const.},
\end{equation}
which we exploit to determine the central charge $c_{\rm eff}$ (see Fig.~\ref{fig:fig4}\,(c) and (d)).
We find a central charge of unity, as expected from a free boson field theory, and that long-range interactions have no effect on the central charge. In the vicinity of the CSB transition, the central charge seems to increase. We attribute this to the presence of a junction where the Haldane line and the U(1) CSB transition meet (c.\,f.~Fig.~\ref{fig:fig1}). We did not check whether long-range interactions alter the universality or the order of the transition analogous to the XXZ anisotropy~\cite{Tzeng2008b,Ueda2008,Hu2011}, and we leave this to future investigations.

\subsection{Ising transition}

We investigate the critical properties of the second-order quantum phase transition between the $\t{Z}_2$ AF and Haldane phases to benchmark our finite-size scaling approach. We compute the squared staggered z-magnetization 
\begin{equation}
    \langle M^2_z \rangle =\frac{1}{N^2}\sum_{i,j}^N (-1)^{i+j} \langle S_i^zS_j^z \rangle \,
    \label{eq:m_stag_z}
\end{equation}
as a function of the anisotropy $D$. Near the critical point $D_c$, the magnetization obeys the finite-size scaling form~\cite{Fisher1972, Binder1987, Cardy1988Book, Koziol2021,Langheld2022}
\begin{equation}
\langle M^2_z \rangle_{D,L} = L^{-2\beta/\nu}f(L^{1/\nu}(D-D_c))\,,
\label{eq:fss}
\end{equation}
which follows from the scaling hypothesis~\cite{Stanley1999}, where $f$ is a fifth-order polynomial function and $\beta$ and $\nu$ are the critical exponents of the order parameter and the correlation length, respectively. We can exploit the finite-size scaling form by performing a multidimensional fit in the anisotropy $D$ and system size $L$ that determines $D_c$, $\nu$, and $\beta$. With proper rescaling of the data with the fit parameters following Eq.~\eqref{eq:fss}, all data points should collapse onto the same curve~\cite{Binder1987,Cardy1988Book, Stanley1999}. We then obtain the critical quantities in the thermodynamic limit by extrapolating their values from finite system sizes. More information on the data-collapse and extrapolation procedure can be found in Appendix~\ref{app:data_collapse}.

For nearest-neighbor interactions ($\alpha=\infty$), we find a critical value of $D_c=-0.31405(5)$, which is in perfect agreement with the literature values in Refs.~\cite{Chen2003,CamposVenuti2006, Tzeng2008a,Ueda2008,Albuquerque2009}. By increasing the strength of the long-range interactions, the $\t{Z}_2$ AF phase stabilizes and the critical line bends towards smaller $D$ values, terminating at \mbox{$D=0$} where the criticality changes to SU(2) CSB as can be inferred from Fig.~\ref{fig:fig1}. As the quantum phase transition from the Haldane to the $\t{Z}_2$ AF phase breaks the Hamiltonian's $\mathbb{Z}_2$ symmetry, we expect a 2D Ising transition with $\nu=1$ and $\beta=1/8$.
\begin{figure}[t]
	\centering
	\includegraphics[width=1.\columnwidth]{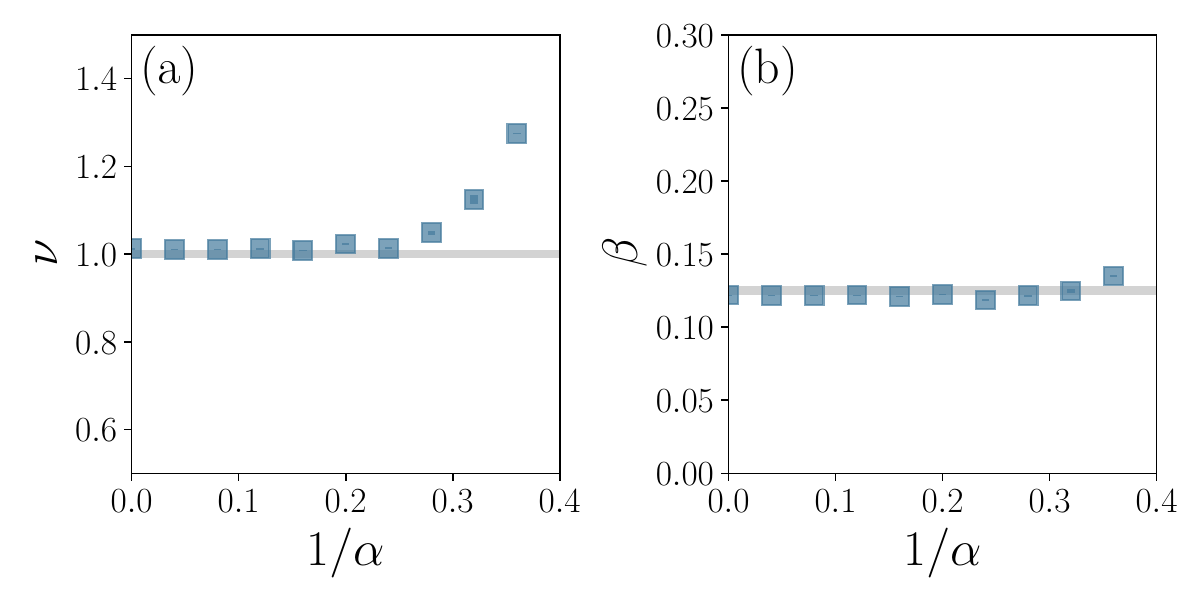}
	\caption{Critical exponents $\nu$ (panel (a)) and $\beta$ (panel (b)) and the corresponding errors---including statistical and systematic uncertainties---from the extrapolation procedure of the Ising transition between the Haldane and $\t{Z}_2$ AF phase from data collapses of the staggered $M_z$ magnetization curves for various decay exponents $\alpha$. Gray lines represent the expected critical values for a 2D Ising transition. The critical exponents are close to the expected values but start to deviate when the critical Ising line approaches the SU(2) CSB critical point.}
	\label{fig:fig5}
\end{figure}
We show the results for the  critical exponents $\nu$ and $\beta$ from the data-collapse and extrapolation procedure in Fig.~\ref{fig:fig5}. The exponents are in excellent agreement with the expectations, although we slightly overestimate (underestimate) the critical exponent $\nu$ ($\beta$). In addition, we see that the presence of the SU(2) CSB transition at $D=0$ spoils the quality of the Ising exponents at smaller $\alpha$ values.

\subsection{Continuous symmetry breaking}

We now turn to the CSB transitions at $D\ge 0$. For values of $D\lesssim 1$, the critical line divides the Haldane phase from the SU(2) CSB phase at $D=0$ and from the U(1) CSB phase for $D\neq 0$. At larger values $D\gtrsim 1$, the transition occurs between the U(1) CSB and large-D phases. In the limit $D\rightarrow \infty$, the CSB phase eventually vanishes for $\alpha>1$ as the $D$-term becomes dominant (c.\,f.~Fig.~\ref{fig:fig1}). Analogously to the Ising transition, we now calculate the staggered transverse magnetization
\begin{equation}
    \langle M^2_{\perp} \rangle =\frac{1}{N^2}\sum_{i,j}^N (-1)^{i+j} \langle S_i^+S_j^- \rangle 
    \label{eq:m_stag_perp}
\end{equation}
as a function of the parameters $\lambda$ and $\alpha$ to probe the critical lines along two different directions. Performing a data collapse of the staggered-magnetization curves $M_\perp$ of different system sizes yields the critical parameters $\lambda_c$ or $\alpha_c$, and exponents $\beta$ and $\nu$ according to the finite-size scaling forms
\begin{equation}
\begin{split}
        \langle M^2_{\perp} \rangle_{\lambda,L} &= L^{-2\beta/\nu}f(L^{1/\nu}(\lambda-\lambda_c))\,, \\
    \langle M^2_{\perp} \rangle_{\alpha,L} &= L^{-2\beta/\nu}f(L^{1/\nu}(\alpha-\alpha_c)) \,.
\end{split}
  \label{eq:fss_forms}
\end{equation}
We again refer the reader to Appendix~\ref{app:data_collapse} for more details on the data-collapse and finite-size extrapolation procedure. Moreover, we complement the MPS data with high-order series expansions using the pCUT+MC approach~\cite{Adelhardt2024} in the perturbation parameter $\lambda=1/D$ about the large-D limit. Using pCUT+MC, we determine the critical line from the closing of the elementary one-quasiparticle (1qp) excitation gap, $\Delta = \omega^{\mathrm{1qp}}(k_c)$, where $k_c=\pi$ is the critical gap momentum. This line agrees very well with the critical line obtained from the data collapse of the MPS data along both the $\lambda$ and $\alpha$ axes, as shown in Fig.~\ref{fig:fig1}. We can determine the critical exponent $z\nu$ associated with the closing of the elementary gap at the critical point $\lambda_c$:
\begin{equation}
    \Delta \sim |\lambda-\lambda_c|^{z\nu}\,.
\end{equation}
The divergence of the associated one-quasiparticle spectral weight, $\mathcal{S}^{1\t{qp}}(k_c)$, defines the critical exponent $\gamma$ through the Fisher scaling relation  $\gamma = (2-\eta)\nu$ and the gap exponent $z\nu$:
\begin{equation}
    \mathcal{S}^{1\t{qp}}(k_c) \sim |\lambda-\lambda_c|^{-(2-z-\eta)\nu}\,.
\end{equation}
See Ref.~\cite{Adelhardt2024} for more information on the pCUT+MC approach itself and the extraction of critical properties. 

In the following, we discuss the phase transition lines of the large-D-to-U(1)-CSB transition and then the Haldane-to-CSB transitions.

\subsubsection{Large-D-to-U(1)-CSB transition}

We first turn our attention to the large-D-to-U(1)-CSB transition at $D\gtrsim 1$, which spans a wide range of different long-range couplings, characterized by the decay exponent $\alpha$. In many long-range spin models, the long-range coupling is a relevant parameter in the renormalization-group sense and can significantly alter the critical behavior of these systems. For instance, in the ferromagnetic transverse-field Ising model and various models with unfrustrated Heisenberg interactions~\cite{Adelhardt2024}, the decay exponent is often parameterized in terms of $\sigma=\alpha-d$ for spatial dimension $d$. This parameter $\sigma$ quantifies the deviation from marginality at $\sigma_\star$ and determines the scaling properties of the long-range interaction under renormalization, thereby controlling its relevance and the resulting critical behavior~\cite{Dutta2001,Defenu2017,Defenu2020}. For $\sigma \ge \sigma_\star$, long-range interactions are irrelevant and the quantum phase transition belongs to the same universality class as the corresponding nearest-neighbor model. Below an upper critical exponent $\sigma \le \sigma_{\t{uc}}$, the transition exhibits Gaussian (long-range mean-field) behavior. Between these two regimes, $\sigma_{\t{uc}} < \sigma < \sigma_\star$, the system displays unconventional long-range criticality with continuously varying critical exponents~\cite{Dutta2001,Defenu2017,Defenu2020}.  For our model in $d=1$, $\sigma_\star$ plays the role of a lower critical exponent, marking the boundary above which a nearest-neighbor critical regime cannot occur. This is a consequence of the Hohenberg-Mermin-Wagner theorem, which rules out such a regime for one-dimensional systems with continuous symmetries~\cite{Pitaevskii1991, Dutta2001, Defenu2017,Defenu2020}.

Using pCUT+MC and MPS, we now probe the critical exponents as a function of the decay exponent to see whether we can replicate a similar behavior as described above. In the following, we use the exponent $\sigma = \alpha - d$ for $d=1$, following standard field-theoretic notation.
\begin{figure}[t]
	\centering
	\includegraphics[width=1.\columnwidth]{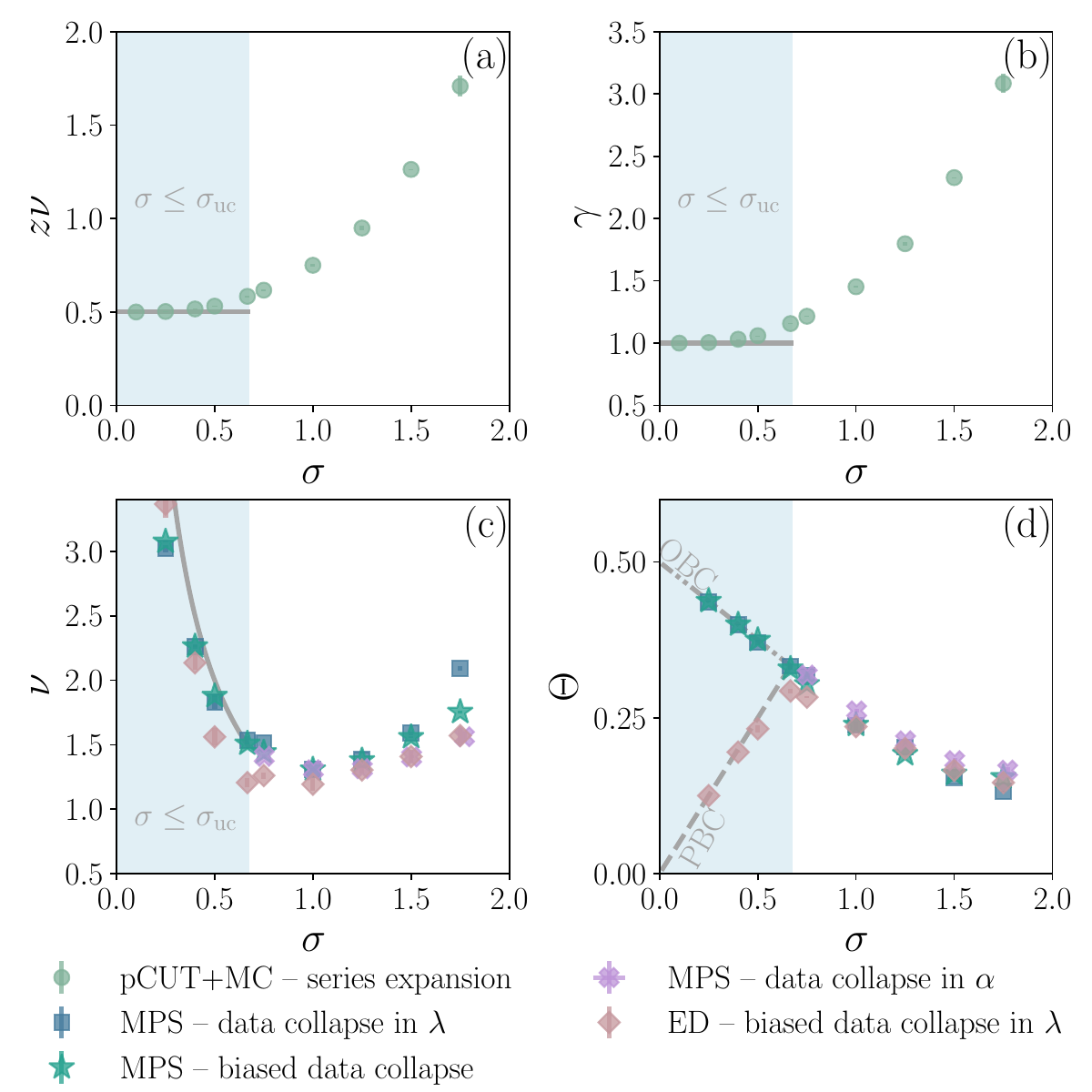}
	\caption{Critical exponents $z\nu$ (a), $\gamma$ (b), $\nu$ (c), and $\Theta$ (d) and their errors---including statistical and systematic uncertainties---obtained from the data-collapse extrapolation procedure  for the large-D-to-U(1)-CSB transition determined from pCUT+MC (panels (a) and (b)), MPS and ED (panels (c) and (d)) as a function of the decay exponent $\sigma = \alpha-d$, where $d=1$. Gray lines indicate the expected long-range mean-field values from the long-range O(2) quantum rotor model above the upper critical dimension \mbox{$\sigma \le \sigma_{\t{uc}}= 2/3$} indicated by the blue shaded region, where $\sigma_{\t{uc}}$ is the corresponding upper critical exponent. Dash-dotted and dashed gray lines indicate the distinct scaling behavior for systems with open (OBC) and periodic boundary conditions (PBC), respectively. Data collapses were performed with respect to $\lambda = 1/D$ (unbiased and biased with the pCUT+MC critical point), as well as with respect to $\sigma$. The exponents show continuously varying behavior for $\sigma \gtrsim 0.5$, while the long-range mean-field behavior is recovered for smaller $\sigma$, which is consistent with $\sigma_{\t{uc}}= 2/3$. The scaling exponent $\Theta$ shows distinct behavior depending on the boundary conditions employed (OBC for MPS and PBC for ED).}
	\label{fig:fig6}
\end{figure}
In Fig.~\ref{fig:fig6}, we show both critical exponents $z\nu$ and $\gamma$ as functions of the decay exponent $\sigma$. For sufficiently small $\sigma \le \sigma_{\mathrm{uc}}$, i.\,e.\,, below the upper critical value, we observe the expected long-range mean-field behavior, with $z\nu \approx 0.5$ and $\gamma \approx 1$; for $\sigma \gtrsim 0.5$, both exponents increase continuously. This behavior is consistent with the upper critical exponent $\sigma_{\mathrm{uc}} = 2/3$ and the divergence of critical exponents predicted for long-range O(N) quantum rotor models~\cite{Dutta2001,Defenu2017,Defenu2020}. Similar behavior has been reported in long-range Heisenberg ladders~\cite{Adelhardt2023,Adelhardt2024,Simon2024Master}, which are described exactly by this effective field theory.

The critical exponent $\nu$ from the data collapse of the MPS magnetization curves following the finite-size scaling forms \eqref{eq:fss_forms} supports the same scenario as the pCUT+MC results. 
The exponent $\nu$ matches the divergent behavior expected in the Gaussian regime at small values of $\sigma$. 
Overall, it shows a similar u-shaped behavior as in long-range Heisenberg ladders~\cite{Adelhardt2024,Simon2024Master,Defenu2017,Defenu2020} as well as the long-range Bose-Hubbard model~\cite{Gupta2024}. 

As it turns out, it is not possible to extract the exponent $\beta$ above the upper critical dimension directly at the critical point for systems with open boundary conditions as used in our MPS approach~\cite{FloresSola2016,Berche2022}.
Instead, we generalize the finite-size scaling forms \eqref{eq:fss_forms} to incorporate a dependence on the boundary conditions of the system. 

Above the upper critical dimension, the conventional finite-size scaling form has to be modified. For periodic boundary conditions, this is described by Q finite-size scaling~\cite{Berche2012,Kenna2013,Kenna2014,Langheld2022}, where the finite-size correlation length scales as $\xi \propto L^\koppa$ with a pseudocritical exponent $\koppa$. Accordingly, we generalize the scaling forms in $\lambda$ as
\begin{align}
     \langle M^2_{\perp} \rangle_{\lambda,L} &= \begin{cases} L^{-2\Theta}f(L^{1/\nu}(\lambda-\lambda_c)) &\t{OBC}\,, \\
     L^{-2\koppa\Theta}f(L^{\koppa/\nu}(\lambda-\lambda_c)) &\t{PBC}\,, 
     \end{cases}
\end{align}
with different dependence for open boundary conditions (OBC) and periodic boundary conditions (PBC), where we introduced the general scaling exponent $\Theta$.
Note, the pseudocritical exponent $\koppa$ is not an independent critical exponent, but a scaling correction exponent that accounts for the modified scaling of the correlation length above the upper critical dimension.
For decay exponents $\sigma > \sigma_{\t{uc}}$, i.\,e.\,, below the upper critical dimension ($d<d_{\t{uc}}$), we have $\koppa=1$, so both scaling forms are identical, and we recover Eqs.~\eqref{eq:fss_forms} where $\Theta=\beta/\nu$.
However, for  $\sigma \le \sigma_{\t{uc}}$ above the upper critical dimension ($d \ge d_{\t{uc}}$), the finite-size scaling forms about the critical point fundamentally differ depending on the boundary conditions of the system under investigation~\cite{Wittmann2014,FloresSola2016,Berche2022}. 

For PBC, the scaling formalism above the upper critical dimension is known as Q finite-size scaling~\cite{Berche2012,Kenna2013,Kenna2014,Langheld2022} with subtle differences between classical and quantum models. For the classical long-range case, the pseudocritical exponent is given by \mbox{$\koppa \equiv d^{\t{cl.}}/d^{\t{cl.}}_{\t{uc}}=d^{\t{cl.}}/2\sigma$}, where we used $d^{\t{cl.}}_{\t{uc}} = 2\sigma$~\cite{Berche2012,Kenna2013,Kenna2014}. 
In contrast, for the quantum long-range case, \mbox{$\koppa=d/d_{\t{uc}}=2d/3\sigma$}, where \mbox{$d_{\t{uc}} = 3\sigma/2$}~\cite{Langheld2022}. In both cases the exponent $\Theta=\beta/\nu$, appearing in the scaling forms, remains unchanged. For OBC, however, the scaling behavior is modified~\cite{FloresSola2016,Berche2022}. 
In Ref.~\cite{FloresSola2016} it was shown that for classical long-range O(N) vector models the exponent becomes \mbox{$\Theta = (d^{\t{cl.}}-\sigma)/2$}. 
Using the quantum-classical mapping $d^{\t{cl.}}=d+z$~\cite{Vojta2003Review,Sachdev2011Book} and $z=\sigma/2$, this yields \mbox{$\Theta = (d-\sigma/2)/2$} for the corresponding quantum model. 
In short, above the upper critical dimension $\sigma\le \sigma_{\t{uc}}$ we expect a distinct scaling behavior for our model depending on the boundary conditions: for OBC, we anticipate \mbox{$\Theta = (1-\sigma/2)/2$}, whereas for PBC, \mbox{$\Theta = \beta/\nu = \sigma/2$} together with a pseudocritical exponent $\koppa=2/3\sigma$. 
Since DMRG typically employs OBC, this prediction implies that critical exponents from such approaches cannot be directly compared to periodic-boundary or field-theoretical predictions without accounting for boundary effects.

To demonstrate these differences in the finite-size scaling above the upper critical dimension, we complement the MPS simulations with OBC by exact diagonalization (ED) calculations using PBC. 
For the ED calculations, we use renormalized long-range couplings obtained via Ewald summation~\cite{Fukui2009,Humeniuk2020,Koziol2021}. 
We apply the same data-collapse and extrapolation procedure as for the MPS data, considering system sizes from $L=6$ to $L=20$. To mitigate the limitations of small accessible system sizes, we perform biased data collapses by fixing the critical parameter to $\lambda_c^{\mathrm{pCUT}}$ obtained from the pCUT+MC approach, which is directly formulated in the thermodynamic limit. 
This reduces the number of free fitting parameters and incorporates an accurate estimate of the transition point, thereby improving the reliability of the extracted critical exponents.

As conjectured, all approaches yield a consistent scaling exponent $\Theta$ for $\sigma>\sigma_{\t{uc}}$ that monotonously decreases from $\sigma=2/3$ to $\sigma = 1.75$ (c.\,f.~Fig~\ref{fig:fig6}). 
Above the upper critical dimension $\sigma\le\sigma_{\t{uc}}$, we observe the boundary condition dependence of the exponent $\Theta$: for OBC, $\Theta$ linearly increases with decreasing $\sigma$, while for PBC it linearly approaches zero. 
Both results agree well with our theoretical predictions ($\Theta = (1/2-\sigma/4)$ for OBC and $\Theta = \sigma/2$ for PBC). 
This clearly demonstrates that boundary conditions are a relevant property in finite-size scaling above the upper critical dimension, rather than a numerical detail or artifact. 
Failure to account for the boundary-dependent scaling may therefore lead to a systematic misidentification of universality classes in long-range systems. 

Note that we also performed biased MPS data collapses with the critical point fixed to the value $\lambda_c^{\mathrm{pCUT}}$; we use these as a consistency check. 
The quantitative agreement between data obtained from pCUT+MC, MPS, and ED, together with their consistency with the predictions of the long-range O(N) quantum rotor model---including the overall qualitative behavior and the location of the upper critical dimension as well as the distinct boundary-dependent scaling behavior above it---provides consistent evidence that the phase transition for sufficiently small decay exponents $\sigma$ is described by a long-range $\phi^4$-theory with O(2) symmetry.

\subsubsection{Haldane-to-CSB transitions}

\begin{figure}[t!]
	\centering
	\includegraphics[width=1.\columnwidth]{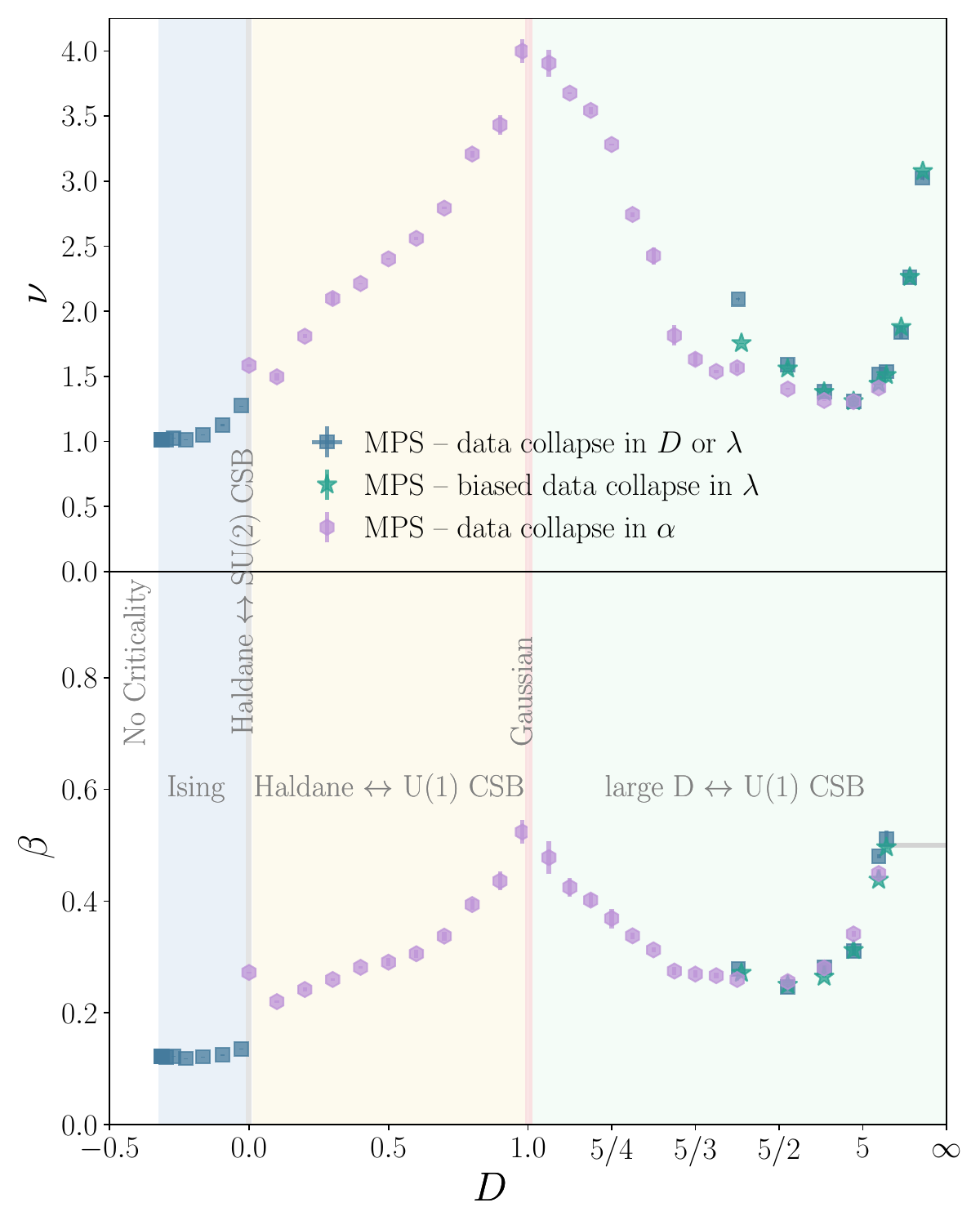}
	\caption{Critical exponents $\nu$ and $\beta$ and their errors---including statistical and systematic uncertainties---obtained from the scaling procedure in the different critical regimes as a function of the anisotropy parameter $D$. The $D$-axis is linear for $D\le 1$, while for $D>1$ it is uniformly spaced in $1/D$. The results are obtained from three different finite-size scaling procedures using MPS: standard data collapse in $D$ and $\lambda$, biased collapse using the critical point $\lambda_c^{\t{pCUT}}$ from the pCUT+MC approach, and a data collapse in the decay exponent $\alpha$.  Shaded background regions mark the different transitions: no criticality ($D < -0.31$, $\t{Z}_2$ AF ground state), the Ising transition between the Haldane and $\t{Z}_2$ AF phases ($-0.31 \le D < 0$), continuous-symmetry-breaking (CSB) transitions from the Haldane phase (SU(2) CSB at $D=0$ and U(1) CSB for $0 < D < 1$), the Gaussian transition near $D \approx 1$ separating the Haldane and large-$D$ phases, and the large-$D$–to–U(1) CSB transition. For $D>1$, the exponents $\nu$ and $\beta$ vary continuously and follow a qualitative u-shaped behavior that is consistent with field-theoretical predictions from the long-range O(2) quantum rotor model~\cite{Adelhardt2024,Defenu2020}. The gray line at large D values ($\beta=0.5$) indicates the expected mean-field value of the exponent $\beta$. Data points obtained from the three approaches largely agree with each other. In the intermediate Haldane-to-U(1)-CSB region ($0 < D < 1$), the system also exhibits unconventional critical behavior with continuously varying exponents as a function of $D$.}
	\label{fig:fig7}
\end{figure}

We now turn our attention to the Haldane-to-CSB transitions. 
To examine the relevant critical line, we show in Fig.~\ref{fig:fig7} the critical exponents $\nu$ and $\beta$ as a function of the anisotropy parameter $D$ over the entire parameter range and indicate the different critical regimes by different background colors.
For $D\lesssim-0.31$, there is no transition as the system has a $\t{Z}_2$ AF ground state for any decay exponent. 
For $-0.31\lesssim D < 0$, the Ising transition between the Haldane and $\t{Z}_2$ AF phase yields critical exponents $\nu\approx 1$ and $\beta\approx 1/8$. 
At $D=0$, a second-order transition separates the Haldane and SU(2) CSB phases, while for $D>0$ the transition is towards U(1) CSB, resulting in an apparent jump of the critical exponents at $D=0$.
We also find that the critical exponents vary continuously as a function of $D$ with a pronounced maximum at $D\approx 1$, where the Gaussian transition terminates in the U(1) CSB critical line. 

For $D > 1$, we again have our large-D-to-U(1)-CSB transition. 
The exponent $\beta$ ($\nu$) continuously decreases from $\beta \approx 0.5$ ($\nu \approx 4.0$) to $\beta \approx 0.25$ ($\nu\approx 1.3$) until $D\approx 5/2$ and then increases continuously. 
The critical exponents are fundamentally controlled by the decay exponent $\sigma$.
When plotting the exponents as a function of the anisotropy $D$, one effectively evaluates them along the line $D_c(\sigma)$.
The observed $D$-dependence of the critical exponents, therefore, reflects the variation of $\sigma$ along the critical line. 
Thus, the described behavior of critical exponents is in line with the expected behavior of the long-range O(2) quantum rotor model~\cite{Dutta2001,Defenu2017,Defenu2020} predicting a u-shaped behavior as a function of $\sigma$~\cite{Defenu2020,Adelhardt2024,Simon2024Master}. 
For sufficiently large anisotropies ($D\gtrsim 7$), the exponents approach their long-range mean-field values. 
The discrepancies in $\nu$ near $D\approx2$ likely reflect the difficulty of accurately determining the critical point due to the very flat critical line in $\lambda$ direction (see Fig.~\ref{fig:fig1}).

For the Haldane-to-U(1)-CSB transition in the range $0<D\lesssim 1$, we find continuously changing critical exponents varying from $\nu \approx 1.5$ to $\approx 4.0$ and from $\beta \approx 0.25$ to $\approx 0.5$. 
The exact origin of this varying critical behavior is not obvious to us.
 
\begin{figure}[t]
	\centering
	\includegraphics[width=1.\columnwidth]{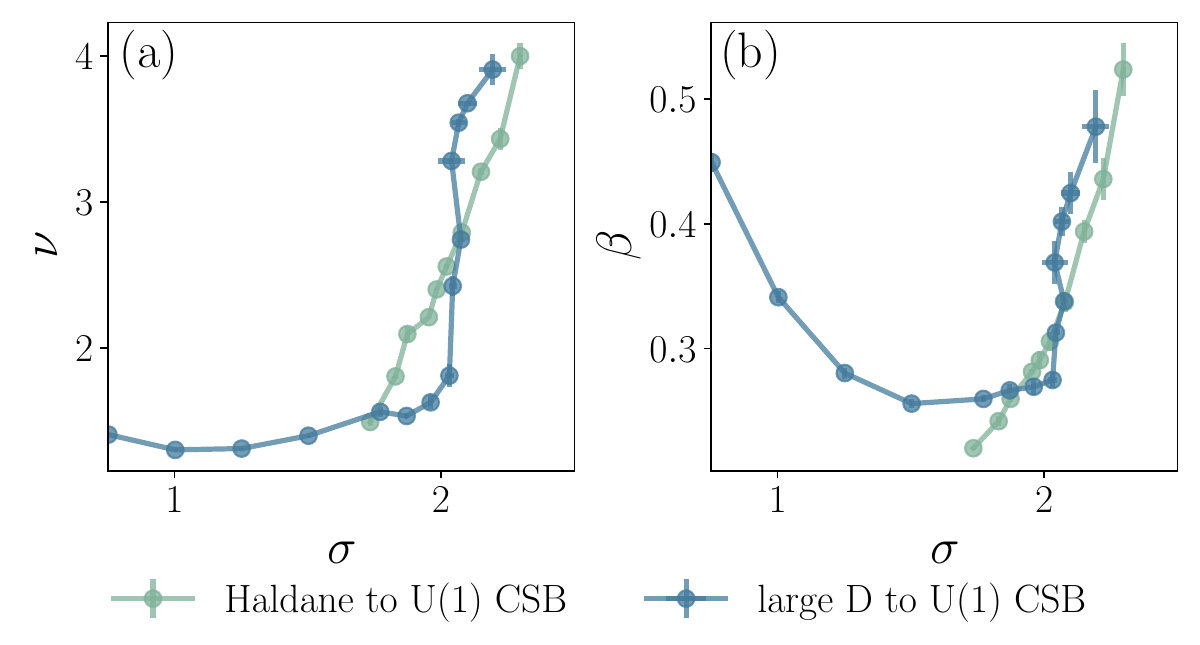}
	\caption{Comparison of the critical exponents $\nu$ (panel (a)) and $\beta$ (panel (b)), including statistical and systematic uncertainties from the scaling analysis, for the Haldane-to-U(1)-CSB and large-D-to-U(1)-CSB transitions as a function of the decay exponent $\sigma=\alpha-d$, where $d=1$. The exponents associated with the large-D-to-U(1)-CSB transition display smoothly varying exponents with a sharp increase at $\sigma\approx 2$, whereas those of the Haldane-to-U(1)-CSB transition increase smoothly around $\sigma \approx 2$.}
	\label{fig:fig8}
\end{figure}

Analogous to the large-D-to-U(1)-CSB transition, one possible interpretation is that the continuously varying exponents arise from an effective ``reparameterization'' of $D_c$ in terms of $\sigma$.
If both critical lines (Haldane-to-U(1)-CSB and large-D-to-U(1)-CSB) were governed by the same underlying effective long-range field theory, where the long-range coupling fixed by $\sigma$ is the relevant parameter in the renormalization-group sense, then the critical exponents should collapse onto a single curve when expressed as functions of the relevant coupling parameter  $\sigma$.
To test this scenario, we plot the critical exponents $\nu$ and $\beta$ for both transitions as functions of $\sigma$ (see Fig.~\ref{fig:fig8}). 
However, the data clearly separate into two distinct branches, demonstrating that the two transitions do not share a common $\sigma$-dependent scaling behavior. 
While the critical exponents of the Haldane-to-U(1)-CSB transition vary almost linearly over a small parameter range around $\sigma\approx2$, the critical exponents of the large-D-to-U(1)-CSB transition vary smoothly for $\sigma \lesssim 2$ in qualitative agreement with the expectation of the long-range O(2) quantum rotor model but additionally show a pronounced, almost jump-like, increase near $\sigma \approx 2$. 
We attribute this jump-like behavior in the large-D-to-U(1)-CSB transition to its proximity to the junction with the Gaussian and Haldane-to-U(1)-CSB transition lines (see Fig.~\ref{fig:fig1}). 
Because the long-range coupling controlled by $\sigma$ is the relevant parameter, while $D$ parameterizes the position along the critical line, the exponents can vary sharply with $\sigma$ but smoothly with $D$.
This behavior is in contrast to the Haldane-to-U(1)-CSB transition, which exhibits smoothly varying exponents as a function of both $\sigma$ and $D$. 
The contrasting behavior suggests an intrinsic difference in the way the two transitions respond to competing critical fluctuations.

If this interpretation is correct, it points to qualitatively distinct underlying mechanisms for the two transitions.
Notably, the Haldane phase constitutes a SPT phase, whereas the large-D phase is topologically trivial. 
Although both transitions involve the breaking of the same U(1) symmetry, the presence of nontrivial topology in the Haldane phase may influence the nature of the critical behavior and could lead to deviations from the conventional Ginzburg-Landau-Wilson paradigm~\cite{Wilson1974, Wilson1983}.  

Related speculations in the context of Heisenberg ladders were previously put forward in Ref.~\cite{Yang2022}, invoking a potential deconfined-quantum-critical scenario between a disordered rung-dimer phase with a finite string-order parameter---just as in the Haldane phase---and an SU(2) CSB phase with conventional order. 
While our results are consistent with such an unconventional scenario, the present analysis does not provide direct evidence.

In further support of the notion that the underlying disordered phases play a crucial role, we compare our critical exponents at the high-symmetry point $D=0$ with those reported for the spin-1/2 Heisenberg chain in Ref.~\cite{Zhao2025}. 
Although the ordered phases have the same symmetry, the transitions originate from qualitatively different disordered phases: the SPT Haldane phase in our spin-one model versus a QLRO phase in the spin-1/2 model.
We find critical exponents \mbox{$\beta=0.2724(13)$} and \mbox{$\nu = 1.584(9)$}, which differ significantly from the values reported in the spin-1/2 model, $\beta = 0.57(2)$ and $\nu = 2.16(1)$~\cite{Zhao2025}. 
Interestingly, the QLRO-to-SU(2)-CSB transition in the spin-1/2 chain is characterized by a dynamical critical exponent $z < 1$~\cite{Laflorencie2005}, while the transition between a QLRO and a U(1) CSB phase is of Berezinskii-Kosterlitz-Thouless type~\cite{Maghrebi2017}, signaling unconventional criticality in these cases as well.

Our exponent estimates are further in agreement with the independent quantum Monte Carlo study in Ref.~\cite{Laflorencie2026} for the same model at $D=0$ finding $\beta\approx 0.3$ and $\nu\approx 1.7$, which appeared after submission of our first manuscript. 
Consistent with our unconventional picture, the study finds evidence for $z<1$ at the Haldane-to-SU(2)-CSB transition, providing independent support for unconventional critical behavior beyond conformal symmetry.

Taken together, these results highlight the multifaceted nature of continuous symmetry breaking in one-dimensional systems and suggest that the nature of the disordered phase can qualitatively affect the resulting critical behavior. 
In this context, our results suggest that the topological nature of the Haldane phase may play an important role for the Haldane-to-CSB transitions, motivating further numerical and field-theoretical studies to clarify their precise nature.

\section{Conclusions and outlook}
\label{sec:conclusion}

In this work, we investigated the spin-one Heisenberg chain with single-ion anisotropy and staggered long-range interactions. Using MPS, we determined the ground-state phase diagram as a function of the long-range decay exponent $\alpha$ and anisotropy $D$, identifying the various phases by computing different spin-spin correlations and the entanglement entropy. Within the CSB phases, long-range interactions give rise to an unconventional $\alpha$-dependent scaling behavior of the logarithmic correction of the entanglement entropy, with distinct behavior for U(1) and SU(2) CSB. In addition, we characterized the critical properties of the second-order quantum phase transitions by analyzing their finite-size behavior. For the topological Gaussian transition, we used the crossing points of the entanglement entropy in different magnetization sectors and the logarithmic scaling at the critical point to determine the critical properties at the quantum phase transition. For transitions involving conventional order ($\t{Z}_2$ AF, SU(2) CSB, and U(1) CSB phases), we used staggered-magnetization curves to perform data collapses and extracted critical properties as a function of the decay exponent $\sigma = \alpha-1$ and the anisotropy $D$. Complementing the MPS results with high-order series expansions using the pCUT+MC approach yielded consistent predictions for the critical line of the large-D-to-U(1)-CSB transition. 

We found that the critical exponents vary continuously with the decay exponent $\sigma$, in agreement with effective quantum field theory predictions for the long-range O(2) quantum rotor model~\cite{Dutta2001,Defenu2017,Defenu2020}. Furthermore, our results demonstrate that the finite-size scaling of critical exponents above the upper critical dimension is strongly influenced by boundary conditions: comparison of MPS results with OBC and ED results with PBC reveals significant differences. To account for this, we extended the finite-size scaling framework for open boundary conditions previously established for classical long-range systems in Ref.~\cite{FloresSola2016} to the quantum setting. To our knowledge, this provides the first quantitative investigations of the influence of boundary conditions on quantum long-range critical scaling above the upper critical dimension complementing a recent work on the finite-size scaling above the upper critical dimension of quantum systems~\cite{Langheld2022}. Our work further highlights that accounting for the correct boundary conditions is crucial to correctly extract critical properties in quantum systems above the upper critical dimension.

Despite mapping out the phase diagram with high precision, several open questions remain. These include the precise properties of the quantum phase transition along the Gaussian line as well as the mechanism underlying the continuously varying critical exponents from the Haldane to the U(1) and SU(2) CSB phases. Addressing these questions in future studies will be essential to better understand and test the unconventional critical behavior in one-dimensional quantum spin systems. 

Beyond these theoretical challenges, our results identify this model as a minimal and versatile playground for studying unconventional quantum criticality. Recent advances in atomic and molecular platforms provide realistic routes toward its implementation. Its defining ingredients---spin-one degrees of freedom, single-ion anisotropy, and tunable long-range interactions---are becoming available in state-of-the-art experiments, including trapped ions~\cite{Cohen2014,Cohen2015,Senko2015}, ultracold atoms~\cite{Chung2021}, and Rydberg atom arrays~\cite{Brechtelsbauer2025, Moegerle2025, Liu2024}. In particular, trapped-ion platforms allow tunable power-law interactions over a wide range of decay exponents~\cite{britton_engineered_2012,katz_floquet_2025}, which is essential for accessing the different critical regimes identified here, while neutral atoms in a cavity have demonstrated tunable interaction potentials in spin-one systems with single-ion anisotropy terms~\cite{periwal_programmable_2021}. Moreover, Rydberg atom platforms have demonstrated that a quantitative finite-size scaling analysis to probe the critical behavior similar to our theoretical approach is already experimentally viable~\cite{Ebadi2021}. Overall, these developments establish a direct connection between theory and experiment, positioning the present model as a concrete setting to study the interplay of long-range interactions, continuous symmetry breaking, and topology and to test the unconventional critical behavior identified in our work.

\textit{Note added---} After submission of our initial manuscript, we became aware of a closely related work in Ref.~\cite{Laflorencie2026}, which appeared almost simultaneously and studies the same model in the absence of single-ion anisotropy using quantum Monte Carlo simulations. Their results are consistent with ours and provide independent support for the unconventional critical behavior at the Haldane-to-CSB transition discussed in this work.

\section*{Data availability}
The raw data is available on Zenodo at \mbox{\hyperlink{https://zenodo.org/records/21953905}{10.5281/zenodo.21953905}}. The code used to generate the numerical results presented in this paper can be made available from P.A. (\href{patrick.adelhardt@fau.de}{patrick.adelhardt@fau.de}) upon reasonable
request.

\section*{Acknowledgment}
Tensor network calculations were performed using the TeNPy library~\cite{Hauschild2018,Hauschild2024} and exact diagonalization calculations with the QuSpin library~\cite{Weinberg2017,Weinberg2019}. P.A. thanks Alexander Schuckert, Anantha Rao, Jan Koziol, Calvin Krämer, and Anja Langheld for fruitful discussions. P.A. gratefully acknowledges A.V.G. for hosting a four-month research stay in Maryland and for a highly stimulating collaboration and welcoming research environment in his group.

P.A. and K.P.S. gratefully acknowledge the support by the Deutsche Forschungsgemeinschaft (DFG, German Research Foundation) -- Project-ID 429529648 -- TRR 306 \mbox{QuCoLiMa} (``Quantum Cooperativity of Light and Matter'') and the Munich Quantum Valley, which is supported by the Bavarian state government with funds from the Hightech Agenda Bayern Plus. P.A. and K.P.S. gratefully acknowledge the scientific support and HPC resources provided by the Erlangen National High Performance Computing Center (NHR@FAU) of the Friedrich-Alexander-Universität Erlangen-Nürnberg (FAU) under the NHR project b177dc (``SELRIQS''). NHR funding is provided by federal and Bavarian state authorities. NHR@FAU hardware is partially funded by the German Research Foundation (DFG) -- 440719683. S.R.M. is supported by the
NSF QLCI (award No. OMA-2120757). A.V.G.~was supported in part by the NSF STAQ program, DoE ASCR Quantum Testbed Pathfinder program (award No.~DE-SC0024220), AFOSR MURI, NSF QLCI (award No.~OMA-2120757), ARL (W911NF-24-2-0107), ONR MURI, DARPA SAVaNT ADVENT, and NQVL:QSTD:Pilot:FTL. A.V.G.~also acknowledges support from the U.S.~Department of Energy, Office of Science, National Quantum Information Science Research Centers, Quantum Systems Accelerator (award No.~DE-SCL0000121) and from the U.S.~Department of Energy, Office of Science, Accelerated Research in Quantum Computing, Fundamental Algorithmic Research toward Quantum Utility (FAR-Qu).

P.A.: Conceptualization, data curation, formal analysis, investigation, methodology, writing---original draft, writing---review and editing. S.R.M.: Supervision, investigation, writing---original draft (introduction and abstract), writing---review and editing. K.P.S.: Supervision, writing---review and editing. A.V.G.: Supervision, writing---review and editing. Following the taxonomy \href{https://credit.niso.org}{CRediT} to categorize the contributions of authors.

\appendix

\section{Convergence of MPS calculations}

In this Appendix, we analyze the convergence of the MPS calculations with respect to the maximum bond dimension $\chi$ and the number of exponential terms $N$ used to approximate the long-range power-law interactions by a sum of exponential terms as introduced in Eq.~\eqref{eq:exp_approx}. 
Convergence with respect to system size is addressed through the finite-size scaling analyses presented in Appendices \ref{app:ee_scaling} and \ref{app:data_collapse}.
To assess convergence, we examine the ground-state energy density $\epsilon_0$ and the entanglement entropy $S_{\rm VN}$ as functions of these two parameters. The convergence of $\epsilon_0$ and $S_{\rm VN}$ provides a stringent check on the underlying MPS ground states used for the finite-size scaling analyses.
\begin{figure}[t]
	\centering
	\includegraphics[width=1.\linewidth]{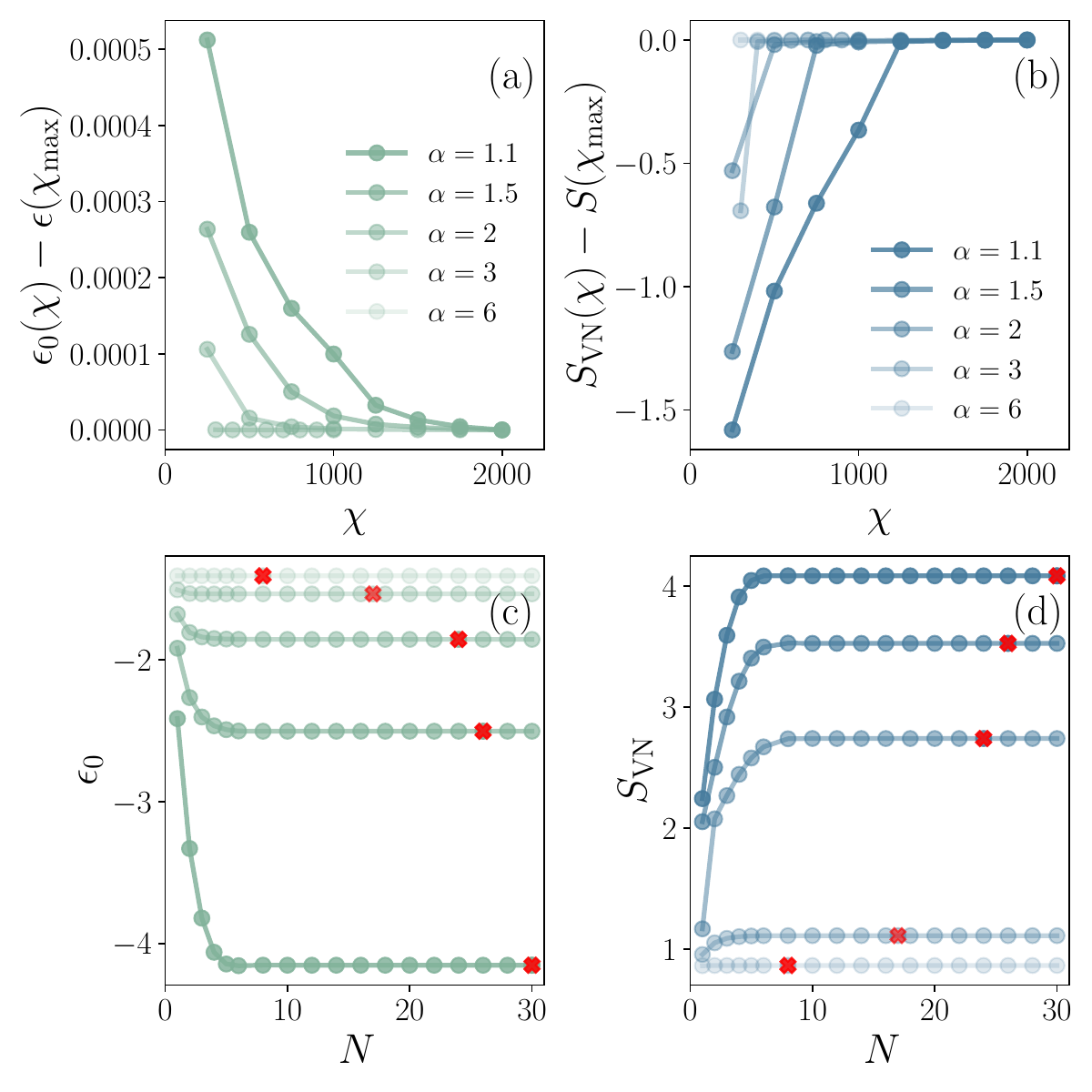}
	\caption{Convergence of the ground-state energy density $\epsilon_0$ and the entanglement entropy $S_{\t{VN}}$ with respect to the bond dimension $\chi$ [panels (a) and (b)] and the number of exponential terms $N$ used to approximate the long-range power-law interaction [panels (c) and (d)] for a chain of length $L=200$ with $D=0$ and various values of the decay exponent $\alpha$. Panels (a) and (b) show the differences $\epsilon_0(\chi)-\epsilon_0(\chi_{\t{max}})$ and $S_{\t{VN}}(\chi)-S_{\t{VN}}(\chi_{\t{max}})$, respectively, relative to the largest bond dimension considered, $\chi_{\t{max}}$, to better illustrate the convergence behavior. Panels (c) and (d) display the dependence on $N$; the red crosses indicate the values of $N$ for which the truncation error satisfies $\varepsilon < 10^{-9}$, corresponding to the choice used throughout our calculations. Both $\epsilon_0$ and $S_{\t{VN}}$ converge more slowly as the decay exponent $\alpha$ decreases.}
	\label{fig:figA1}
\end{figure}
In Fig.~\ref{fig:figA1}, we show the data for a chain with size $L=200$, anisotropy $D=0$ and various decay exponents $\alpha$ between $\alpha=1.1$ and $\alpha=\infty$. 
We choose representative data at $D=0$ where the entanglement is largest across the entire phase diagram and increases substantially as the decay exponent $\alpha$ decreases, making this the most demanding regime for the MPS calculations. We therefore use $D=0$ as a worst-case benchmark for the convergence of the MPS calculations. 
The simulation parameters employed throughout this work---a maximum bond dimension of $\chi_{\t{max}}=2000$ at $D=0$ in the SU(2)-broken CSB phase and $\chi_{\t{max}}=500$ and $\chi_{\t{max}}=1000$ elsewhere, together with a variable number of exponential terms $N$ chosen such that the approximation error satisfies $\varepsilon < 10^{-9}$---yield well-converged results even for the smallest values of $\alpha$ considered. 
These convergence tests indicate that the numerical accuracy achieved is sufficient for the analyses presented in the main text.

\section{Scaling of the entanglement entropy}
\label{app:ee_scaling}

In this Appendix, we present the procedure for determining the scaling behavior in more detail and also discuss the constant scaling term $c$ as a function of the decay exponent $\alpha$. The spin-one Heisenberg chain exhibits a logarithmic scaling of the entanglement entropy following Eq.~\eqref{eq:area_law}, where the leading area-law term becomes trivially a constant in one-dimensional systems. In CSB phases, the low-lying tower-of-states structure of the entanglement spectrum leads to an additive logarithmic correction to the area-law scaling~\cite{Laflorencie2016, Frerot2017, Zhao2025}. Thus, we use a logarithmic function,
\begin{equation}
    S_{\t{VN}}(L) = b \ln(L) + c\,,
\end{equation}
to perform a least-squares fit to the entanglement entropy as a function of the system size $L$ with $b$ and $c$ as free fit parameters.
\begin{figure}[t]
	\centering
	\includegraphics[width=1.\linewidth]{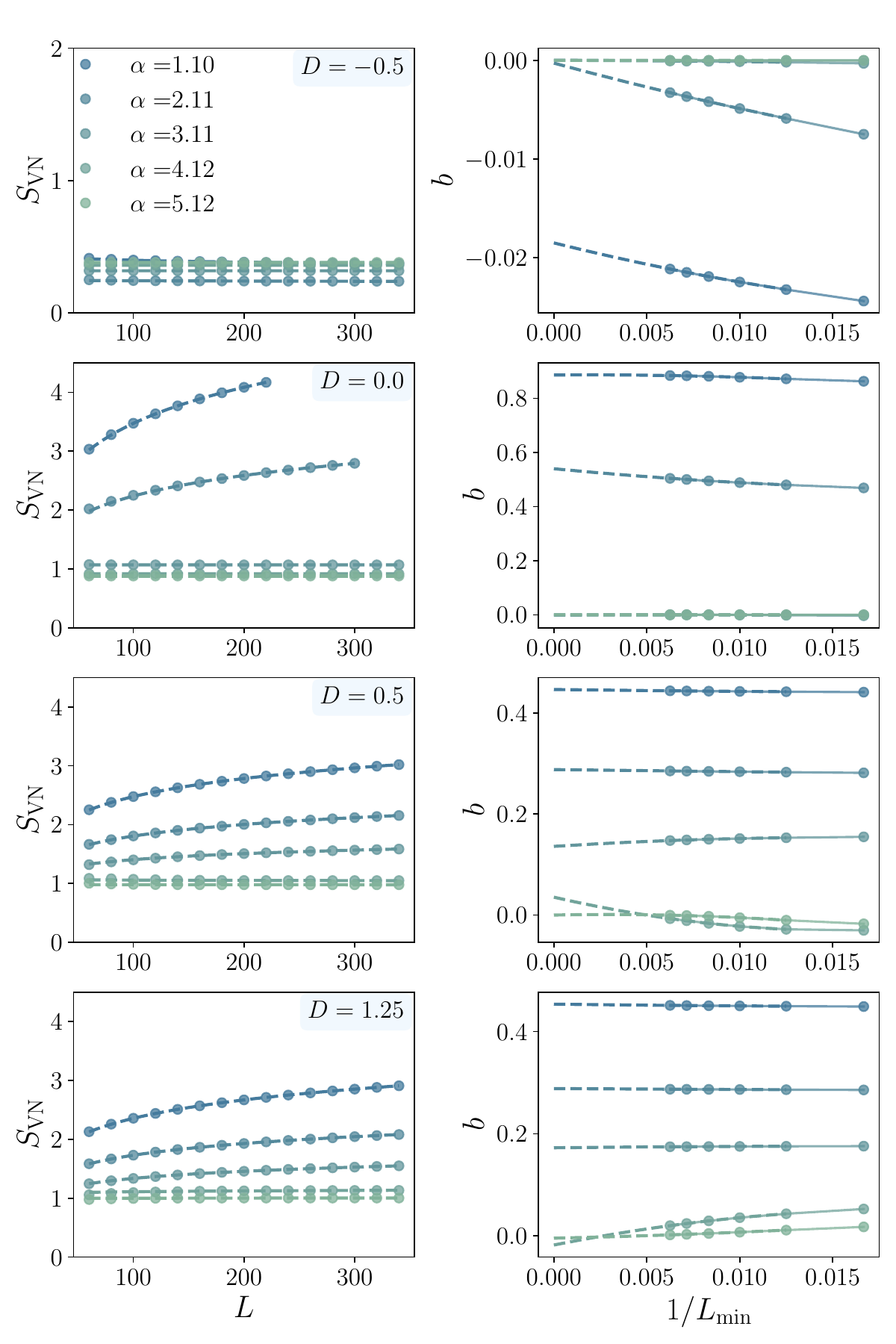}
	\caption{Scaling behavior of the entanglement entropy $S_{\t{VN}}$ for $D \in \{-0.5, 0, 0.5, 1.25\}$ at various decay exponents $\alpha$. \textit{Left column:} Entanglement entropy as a function of system size $L$, fitted to $S_{\t{VN}} = b \ln(L) + c$ (dashed lines). \textit{Right column:} Logarithmic coefficients $b$ plotted versus $1/L_{\min}$, where $L_{\min}$ is the lower bound of a `sliding' fitting window. Each window contains $N=10$ data points to which the logarithmic fit is applied, and the window is shifted sequentially to larger system sizes. Dashed lines show linear fits of $b(1/L_{\min})$ used to extrapolate the thermodynamic value $b(1/L_{\min}=0)$.}
	\label{fig:figA2}
\end{figure}
In the left column of Fig.~\ref{fig:figA2} we show the entanglement entropy as it increases with the system size $L$ and the logarithmic fit function for selected values of the decay exponent $\alpha$ for $D\in\{-0.5,0,0.5,1.25\}$. Note that for $\alpha=1.1$ at $D=0$, the data points are limited to $L\le 220$. Due to enhanced entanglement in the SU(2) CSB phase, we cannot push to the maximal system size ($L=340$) for small $\alpha$-values with a bond dimension limited to $\chi_{\t{max}}=2000$. Until $\alpha\approx 1.85$ the maximum system size $L=340$ is accessible but steadily decreases to $L=220$ at $\alpha=1.1$.

We aim to account for finite-size effects by successively performing least-squares fits in a `sliding window' with a fixed number of system sizes. Thus, we slide the window with the initial set of system sizes \mbox{$L=\{60, 80, 100, 120, \dots, 240\}$} to the final one \mbox{$L=\{160, 180, 200, 220, \dots, 340\}$}, i.\,e.\,, from $L_{\t{min}}=60$ to $L_{\t{min}}=160$ resulting in six data points for the logarithmic coefficient $b$. In the right column of Fig.~\ref{fig:figA2}, we show the logarithmic coefficients $b(1/L_{\t{min}})$ as a function of the minimal system size of the `sliding window'. To extract the bulk value, we use a linear fit function
\begin{equation}
    f(1/L_{\t{min}}) = b_{\infty} + u\,(1/L_{\t{min}})
\end{equation}
in the range $L_{\t{min}}=[80, 160]$, where $b_{\infty}$ and $u$ are the free fit parameters. The coefficient of the logarithmic correction in the thermodynamic limit is then given by $f(0)=b$ with the fitting error $\sigma_{b}$.

\begin{figure}[t]
	\centering
	\includegraphics[width=1.\linewidth]{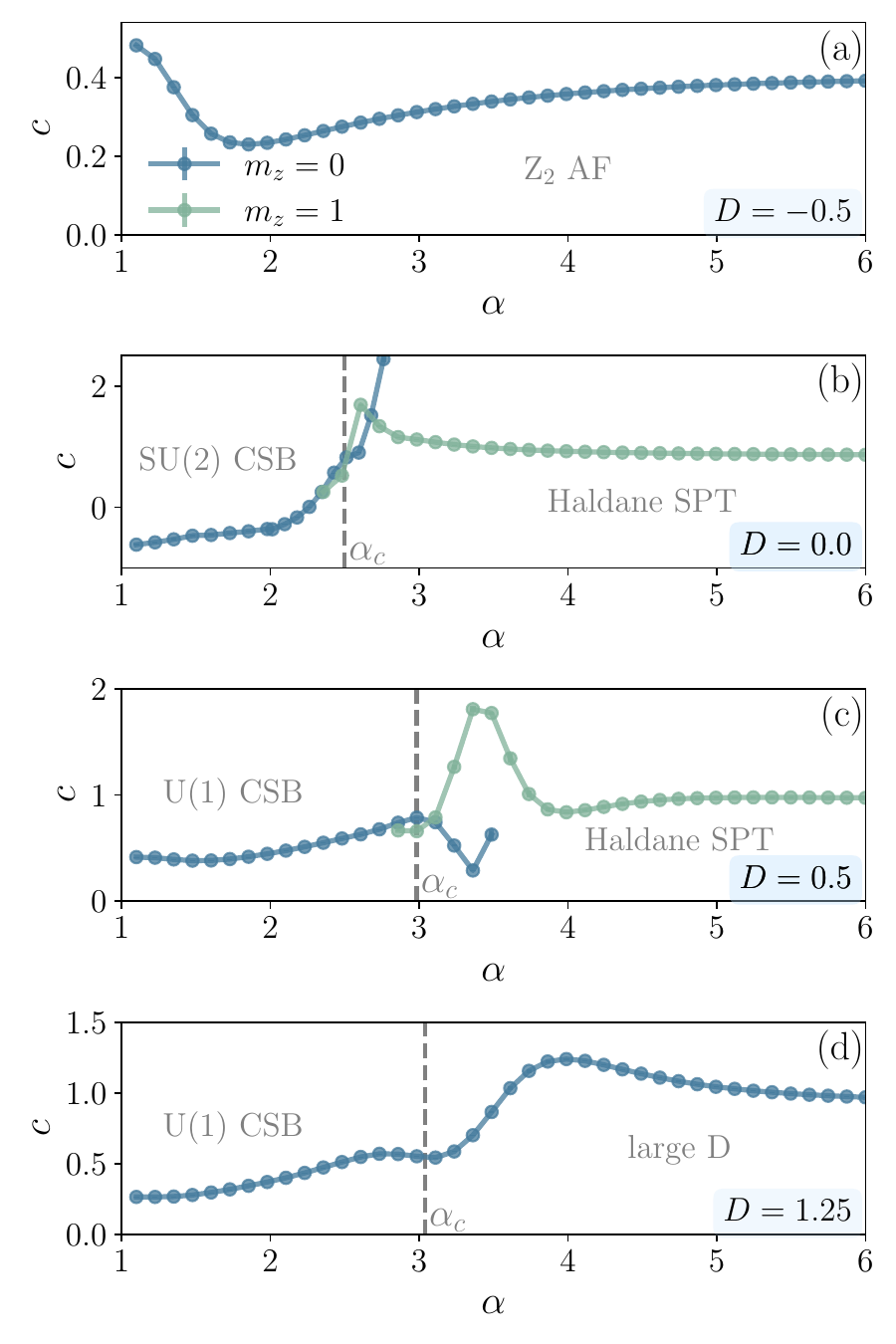}
	\caption{Constant term $c$ of the entanglement entropy as a function of the decay exponent $\alpha$ for \mbox{$D = -0.5$, $0$, $0.5$, and $1.25$} in panels (a), (b), (c), and (d), respectively. Data points include errorbars from the extrapolation procedure. The ground state in the $m_z = 0$ sector is used to compute the entanglement entropy, except in the Haldane phase, where the $m_z = 1$ ground state is taken to avoid the residual $\ln 2$ contribution from edge-state entanglement. Fitting the entanglement entropy to $S_{\mathrm{VN}} = b \ln L + c$ yields the constant $c$, which displays a non-trivial behavior close to the critical points $\alpha_c$. Gray dashed lines indicate critical points in the thermodynamic limit determined by the data-collapse and extrapolation procedure in Appendix~\ref{app:data_collapse}.}
	\label{fig:figA3}
\end{figure}

In the main text we discussed the logarithmic coefficient $b$ as a function of the decay exponent $\alpha$ from the expected entanglement entropy
\begin{equation}
    S_{\t{VN}} = b \ln(L)  + c\,.
\end{equation}
Here, we analogously discuss the behavior of the constant term $c$ as a function of the decay exponent $\alpha$ as shown in Fig.~\ref{fig:figA3}. The constant is obtained from the same fits used in the main text. In contrast to the logarithmic coefficient $b$, the constant term $c$ is, in general, not universal and typically more prone to finite-size effects. Consequently, the results should be interpreted with more caution, especially the behavior close to the critical point.
In the Z$_2$ AF phase at $D=-0.5$ (panel (a)), the constant term varies smoothly, decreasing continuously from $\alpha=6$ to $\alpha\approx2$. For $\alpha\lesssim 2$, $c$ increases again. This behavior coincides with the regime where the logarithmic coefficient $b$ becomes negative. In this regime, the entanglement entropy slightly decreases with increasing system size, which leads to a negative logarithmic slope in the fit and a compensating increase of the constant term $c$. For $D=0$ in panel (b), the constant term is positive and approximately constant in the Haldane phase ($\alpha\gtrsim 2.5$), while it becomes negative in the SU(2) CSB phase ($\alpha\lesssim 2.5$). Near the critical point, $c$ increases from both sides and exhibits a pronounced peak. Such peaks are expected close to the transition where scaling corrections are strongest and the simple logarithmic form provides only an approximate description of the finite-size data. For $D=0.5$ in panel (c), the constant remains positive and approximately constant in both the Haldane and U(1) CSB phases. In addition, a pronounced peak in $c$ appears on the Haldane side of the transition ($\alpha_c\approx3$). Finally, for $D=1.25$ in panel (d), the behavior across the large-D and U(1) CSB phases is qualitatively similar to the previous cases. The constant increases on the large-D side of the transition, reaches a maximum around $\alpha\approx4$, and slowly decreases for larger $\alpha$ values.

\section{Data-collapse and extrapolation of critical properties}
\label{app:data_collapse}

In this Appendix, we present the data-collapse and extrapolation procedure which we apply to extract the critical properties of the second-order transition lines in the phase diagram to high accuracy. Extracting critical properties---most notably critical exponents---is a very challenging task. A very accurate and proven approach is to use the finite-size scaling of observables and perform a data collapse exploiting the universal critical behavior about second-order quantum phase transitions~\cite{Fisher1972, Binder1987, Cardy1988Book, Stanley1999, Koziol2021, Langheld2022}. However, systems governed by long-range interactions can give rise to increased entanglement and long-range correlations leading to stronger finite-size effects, which makes the determination of critical properties even more challenging.

To get accurate results, we apply a two-step procedure. First, we perform several data collapses for different sets of system sizes, which allows us to extract the critical point and exponents. Second, we analyze the behavior of the critical point and exponents as a function of the inverse system size and extrapolate the data to the thermodynamic limit. In the following, we describe these two steps in more detail.

\begin{figure*}[t]
	\centering
	\includegraphics[width=1.\linewidth]{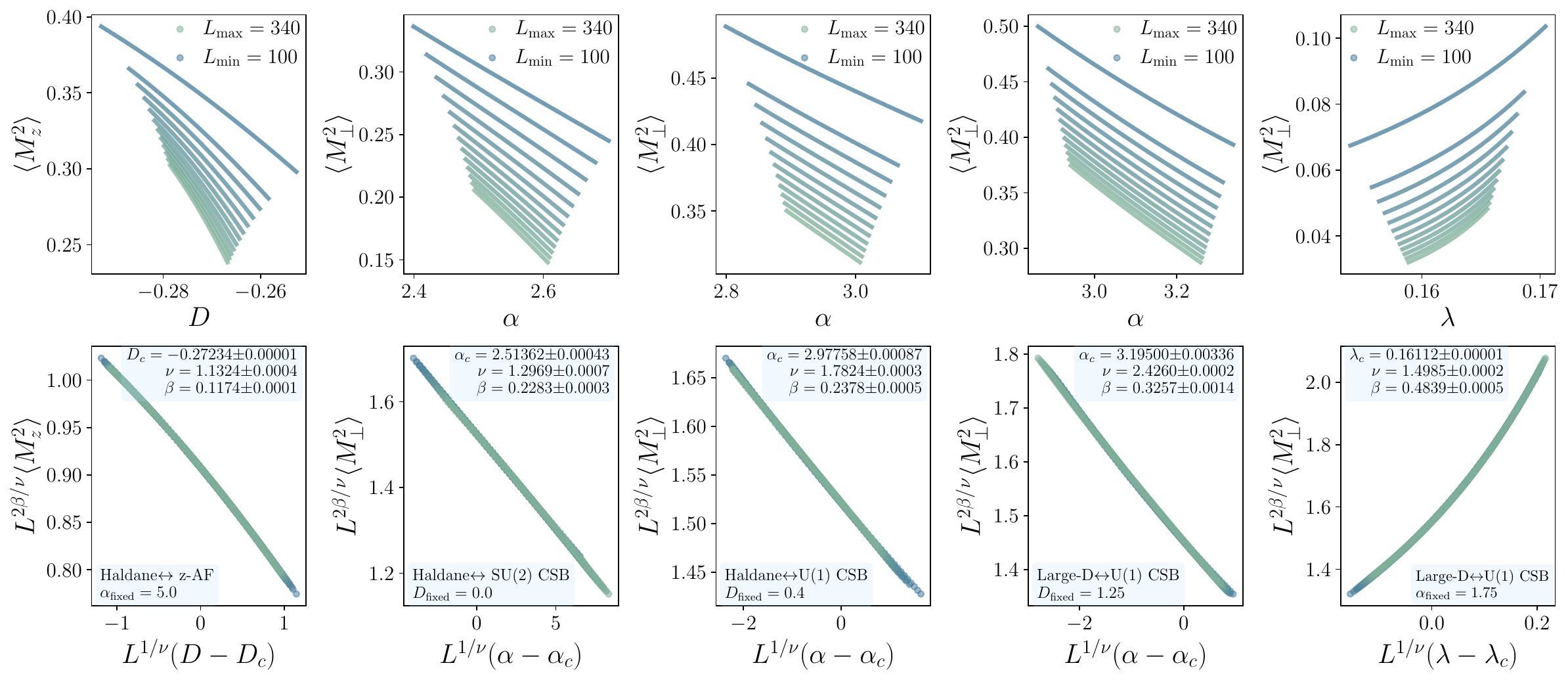}
	\caption{Representative data collapses probing the Ising, SU(2), and U(1) continuous symmetry breaking (CSB) transitions at values $\alpha=5$, $D\in\{0,0.4,1.25\}$, and $\alpha=1.75$. \textit{Top row:} Staggered magnetization curves of $\langle M^2_z \rangle$ for the Ising transition and $\langle M^2_{\perp} \rangle$ for the CSB transitions between $L_{\t{min}}=100$ and $L_{\t{max}}=340$ versus the tuning parameters $D$, $\alpha$, and $\lambda=1/D$. \textit{Bottom row:} The data collapses associated to the magnetization curves in the top row yielding the critical point ($D_c$, $\alpha_c$, or $\lambda_c$) and the corresponding critical exponents $\beta$ and $\nu$ of the second-order quantum phase transitions.}
	\label{fig:figA4}
\end{figure*}
We calculate the staggered-magnetization curves for \mbox{$L=\{60, 100, 140, 160, 180, 200, \dots, 340\}$} in the vicinity of all second-order phase transitions (except the topological Gaussian transition). We use the following finite-size scaling forms 
\begin{equation}
\begin{split}
 \langle M^2_z \rangle_{D,L} &= L^{-2\beta/\nu}f(L^{1/\nu}(D-D_c))\,, \\
 \langle M^2_{\perp} \rangle_{\lambda,L} &= L^{-2\beta/\nu}f(L^{1/\nu}(\lambda-\lambda_c))\,, \\
\langle M^2_{\perp} \rangle_{\alpha,L} &= L^{-2\beta/\nu}f(L^{1/\nu}(\alpha-\alpha_c)) \,,
\end{split}
\label{eq:fss_forms_ext}
\end{equation}
where $f$ is a fifth-order polynomial function. The critical points $D_c$, $\lambda_c$ or $\alpha_c$ and critical exponents $\beta$ and $\nu$ are additional free parameters that we determine by a multi-dimensional least-squares fit in the parameters $L$ and $D$, $\lambda=1/D$, or $\alpha$. For the Ising transition, we use the staggered z-magnetization $M_z$ \eqref{eq:m_stag_z}, while for the transitions toward the CSB phases, we use the staggered transverse magnetization $M_{\perp}$ \eqref{eq:m_stag_perp}. Depending on the critical line, we apply a data collapse in the direction along $\alpha$ for the CSB transitions or in the opposite direction along $D$ for the Ising transition and along $\lambda$ for large $D$ values towards the U(1) CSB transition. Moreover, we perform biased data collapses in $\lambda$, where we fix the critical point $\lambda_c$ to the pCUT+MC value $\lambda_c^{\mathrm{pCUT}}$, which is determined independently using the high-order series-expansion approach. This reduces the number of free fit parameters to two, $\beta$ and $\nu$, serving as a biased fit to check the consistency of our results.

Above the upper critical dimension $d \ge d_{\t{uc}}$---or likewise below the upper critical exponents $\sigma \le \sigma_{\t{uc}}$, where we use an alternative decay exponent $\sigma=\alpha-d$ with $d$ being the dimension of the system---the above finite-size scaling forms must be modified. Interestingly, the finite-size scaling forms differ depending on the boundary conditions. Hence, we use the following more general finite-size scaling forms:
\begin{align}
     \langle M^2_{\perp} \rangle_{\lambda,L} &= \begin{cases} L^{-2\Theta}f(L^{1/\nu}(\lambda-\lambda_c)) &\t{OBC}\,, \\
     L^{-2\koppa\Theta}f(L^{\koppa/\nu}(\lambda-\lambda_c)) &\t{PBC}\,, 
     \end{cases}
\end{align}
for open boundary conditions (OBC) and periodic boundary conditions (PBC), respectively.  We introduce the general scaling exponent $\Theta$ and the pseudocritical exponent $\koppa$. For $\sigma \ge \sigma_{\t{uc}}$, we have $\koppa=1$, both scaling forms are identical, and we recover  Eqs.~\eqref{eq:fss_forms_ext} where we introduce the critical exponent $\Theta=\beta/\nu$.   However, if  $\sigma \le \sigma_{\t{uc}}$, we have $\koppa= 2/3\sigma$ and $\Theta=\beta/\nu$ for PBC, and $\Theta=(d-\sigma/2)/2$ for OBC, which is now independent of $\beta$ and $\nu$. To test this behavior, we performed the data collapses using $\Theta$ as a free parameter.

After performing least-squares fits to the MPS data points with critical exponents (and critical point) as free parameters, the established approach is to collapse the magnetization curves of different system sizes onto a single curve using the final fit parameters and proper rescaling according to the finite-size scaling forms \eqref{eq:fss_forms_ext}. In Fig.~\ref{fig:figA4}, we show the magnetization curves for different system sizes and the associated data collapse representative for different transitions. As we can see, the data points perfectly collapse onto a single curve as expected at second-order quantum phase transitions. To get a reasonable error estimate of the critical values it is not sufficient to only consider the uncertainty from the least-squares fit, as we are limited to small system sizes. In reality, corrections to the dominant power-law scaling are to be expected, and long-range interactions further amplify finite-size effects. 

To systematically account for finite-size effects, we perform multiple data collapses using a `sliding window' with several system sizes. 
The variation between the extrapolations obtained from different sliding windows is interpreted as an estimate of the residual systematic uncertainty arising from finite-size effects. 
A pronounced dependence on the chosen window indicates remaining finite-size effects not captured by the scaling ansatz and leads to an increased systematic uncertainty in the final estimate.
We fix the number of system sizes for the data collapse to eight and `slide' the window from \mbox{$L=\{60, 100, 140, 160, \dots, 240\}$} to \mbox{$L=\{200, 220, 240, 260, \dots, 340\}$}, i.\,e.\,, from $L_{\t{min}}=60$ to $L_{\t{min}}=200$ resulting in six data points.
\begin{figure*}[t]
	\centering
	\includegraphics[width=1.\linewidth]{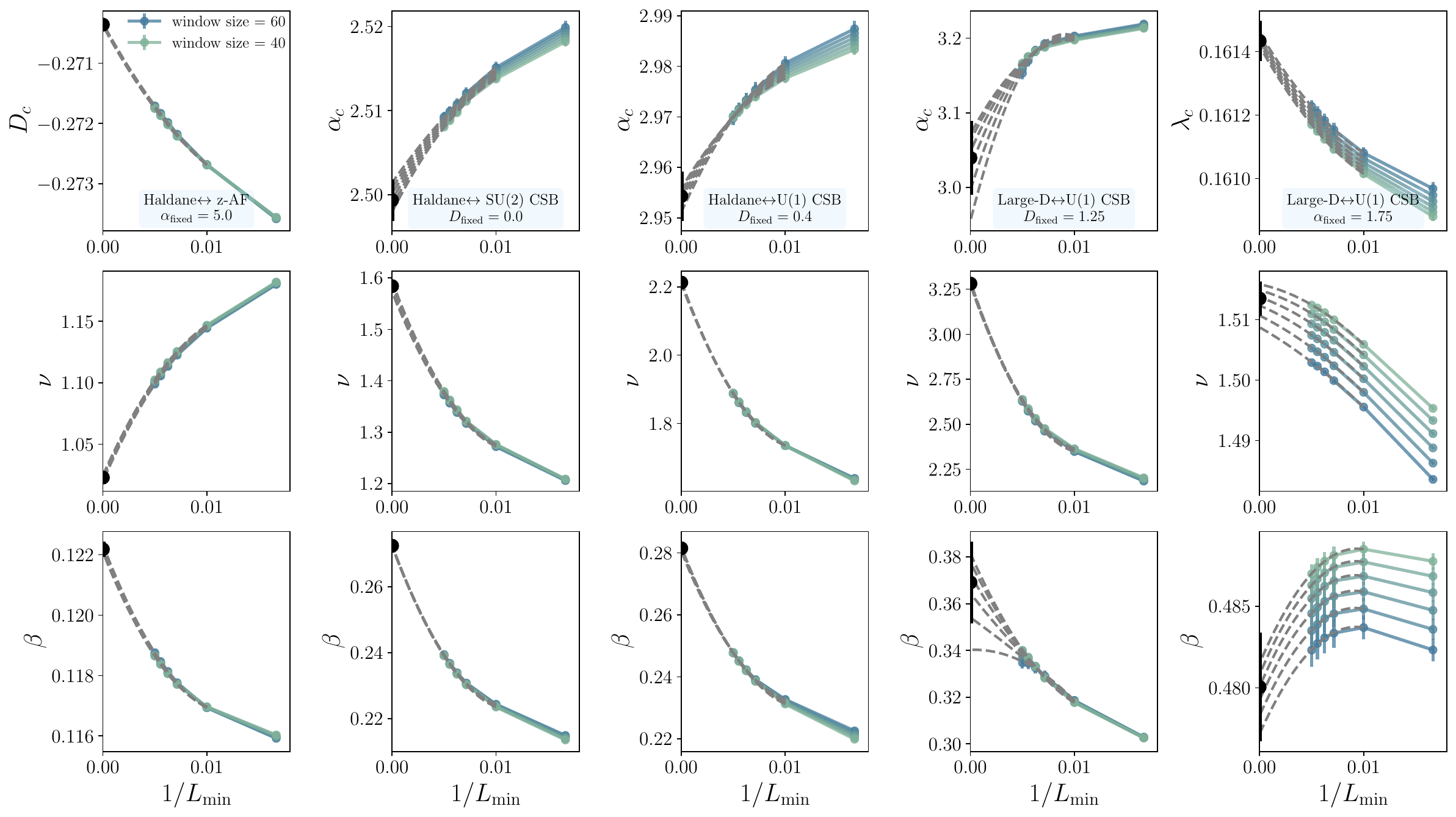}
	\caption{Scaling of the critical points $D_c$, $\alpha_c$, and $\lambda_c = 1/D_c$ (first row), the critical exponents $\nu$ (second row), and $\beta$ (third row) obtained from data collapses at $\alpha = 5$, $D \in \{0, 0.4, 1.25\}$, and $\alpha = 1.75$, plotted versus the inverse system size $1/L_{\min}$. The minimal system size $L_{\min}$ corresponds to the lower bound of a sliding window used for the collapses, with each window containing eight system sizes and shifted sequentially to larger system sizes. Finite-size behavior is shown for data window sizes between 40--60 data points. A quadratic function $f(1/L_{\min}) = \kappa + u\,(1/L_{\min}) + v\,(1/L_{\min})^2$ is fitted to the last five data points (gray dashed lines) to estimate the bulk value $f(1/L_{\min}=0) = \kappa$. The black data points at $1/L_{\min}=0$ represent the weighted average of the critical quantity in the thermodynamic limit including statistical and systematic uncertainties.}
	\label{fig:figA5}
\end{figure*}
In Fig.~\ref{fig:figA5}, we show the behavior of the three critical quantities as a function of the minimal system size of the `sliding' window. We use a quadratic fit function
\begin{equation}
    f(1/L_{\t{min}}) = \kappa + u\,(1/L_{\t{min}})+ v\,(1/L_{\t{min}})^{2}
\end{equation}
to extrapolate their values to the thermodynamic limit, which is given by $f(0)=\kappa$ with the associated fitting error $\sigma_{\kappa}$. Here, the parameter $\kappa$ represents the critical point ($D_c$, $\alpha_c$, and $\lambda_c$) or a critical exponent ($\beta$ and $\nu$). We use a least-squares fit in the range $[100,200]$. 

To further improve the error estimate, we repeat the data-collapse and extrapolation procedure with $N=6$ different data windows where the number of data points ranges between $40-60$ per system size. We obtain several estimates for the bulk values $\kappa_i$ and calculate the mean and associated spread with
\begin{align}
    \bar{\kappa} &= \frac{1}{N} \sum_{i=1}^N \kappa_i\,, \\
    \sigma_\text{spread} &= \sqrt{\frac{1}{N-1} \sum_{i=1}^N (\kappa_i - \bar{\kappa})^2}\,.
\end{align}
As the individual quadratic fits to the different data windows lead to different uncertainties $\sigma_{\kappa,i}$ where $\kappa_i \pm \sigma_{\kappa,i}$, it is better to compute the weighted average 
\begin{equation}
\bar{\kappa}_\text{w} = \frac{\sum_{i=1}^N w_i \kappa_i}{\sum_{i=1}^N w_i}\,, \quad \sigma_\text{w} = \frac{1}{\sqrt{\sum_{i=1}^N w_i}}\,,
\end{equation}
where we define the weights as $w_i = 1/\sigma_{\kappa,i}^2$. The weighted average minimizes the variance under the assumption that the individual $\kappa_i$ values are independent measurements governed by their respective uncertainties.

To test the consistency of the weighted average with the reported errors, we use the (reduced) chi-squared value defined as
\begin{equation}
\chi^2 = \sum_{i=1}^N \frac{(\kappa_i - \bar{\kappa}_\text{w})^2}{\sigma_{\kappa,i}^2}, \quad
\chi^2_\text{red} = \frac{\chi^2}{N-1}\,.
\end{equation}
If $\chi^2_{\mathrm{red}} \le 1$, the scatter among the $\kappa_i$ values is compatible with the fitting uncertainties. When $\chi^2_{\mathrm{red}} > 1$, the observed fluctuations exceed those expected from the errors $\sigma_{\kappa,i}$, indicating that the statistical uncertainties must be inflated. We therefore obtain a statistical error
\begin{equation}
    \sigma_\text{stat} = \sigma_\text{w} \sqrt{\max(\chi^2_\text{red}, 1)}\,.
\end{equation}
In addition to the statistical uncertainty, varying the window size introduces correlations among the data and leads to a systematic spread that is not captured by the individual fit errors. We therefore interpret
\begin{equation}
    \sigma_{\mathrm{syst}} = \sigma_{\mathrm{spread}}
\end{equation}
as an estimate of the residual systematic uncertainty associated with the extrapolation procedure. 
We combine this systematic error with the previous statistical error to a total error estimate 
\begin{equation}
    \sigma_\text{total} = \sqrt{\sigma_\text{stat}^2 + \sigma_\text{syst}^2}\,,
\end{equation}
for our final estimation of the critical quantities given by $\kappa_{\t{w}}\pm \sigma_{\t{total}}$. 
The systematic contribution can significantly increase the total error bar when the extrapolated values show a strong dependence on the chosen sliding window.
In Fig.~\ref{fig:figA5}, we depict the extrapolations and the final estimates for the critical point and the critical exponents $\nu$ and $\beta$. 
The finite-size behavior is well-behaved and allows for decent fits to the quadratic fit function. 
We see that the critical exponents for the Ising transition $\nu = 1.0228(25)$  and $\beta= 0.12218(26)$ at $\alpha=5$ agree closely with the expected values in literature $\nu=1$ and $\beta=0.125$.

\newpage
\bibliography{bibliography}

\end{document}